\documentclass[twocolumn,trackchanges]{aastex7}

\usepackage{amsmath}
\begin{document}
	
\title{A Disk-Originated 329-day Quasi-Periodic Oscillation in the Seyfert 1 Galaxy J1626+5120}

\correspondingauthor{Litao Zhu; Zhongxiang Wang}

\author[]{Litao Zhu}
\affiliation{Department of Astronomy, School of Physics and Astronomy, Key Laboratory of Astroparticle Physics of Yunnan Province, Yunnan, China}
\email{zhulitao@mail.ynu.edu.cn}

\author[orcid=0000-0003-1984-3852]{Zhongxiang Wang}
\affiliation{Department of Astronomy, School of Physics and Astronomy, Key Laboratory of Astroparticle Physics of Yunnan Province, Yunnan, China}
\email{wangzx20@ynu.edu.cn}

\author[]{Dong Zheng}
\affiliation{Department of Astronomy, School of Physics and Astronomy, Key Laboratory of Astroparticle Physics of Yunnan Province, Yunnan, China} 
\email{zhengdong@mail.ynu.edu.cn}

\author[0000-0002-9331-4388]{Alok C. Gupta}
\affiliation{Aryabhatta Research Institute of Observational Sciences (ARIES), Manora Peak, Nainital-263001, India}
\email{acgupta30@gmail.com}

\author[]{Ju-Jia Zhang}
\affiliation{Yunnan Observatories, Chinese Academy of Sciences, Kunming 650216, China}
\affiliation{Key Laboratory for the Structure and Evolution of Celestial Objects, Chinese Academy of Sciences, Kunming 650216, China}
\email{jujia@ynao.ac.cn}

\begin{abstract}
	The Seyfert 1 galaxy J1626+5120 is estimated to host a $10^8 M_{\odot}$
	black hole (BH) accreting at Eddington ratio 
	$\dot{m}_{\text{Edd}} \approx 0.043$. Its long-term multi-band light 
	curve data show flicker-like variations, but in a well-sampled $g$-band 
	light curve, we are able to determine a $\simeq 329$\,d
	quasi-periodic oscillation (QPO) at a $\sim$4.53$\sigma$ significance.
	Six optical 
	spectra were obtained for the source, three of which were taken by us.
	The spectra show that the variations were mainly because of flux changes
	blueward of 4000\,\AA. We also analyze X-ray and ultraviolet (UV) data
	obtained with {\it the Neil Gehrels Swift Observatory (Swift)}, which 
	targeted the source in the past two years. X-ray and UV emissions of the
	source show variations correlated with optical. Time lags of four UV
	bands and four optical bands are determined with respect to the X-ray
emission, which are consistent with a continuum reprocessing disk model.
	These properties point out a disk origin for the QPO, likely due to
        Lense-Thirring (LT) precession of the accretion flow at $\sim$20
	gravitational radii 
	of the BH. This QPO could be a key case linking
	sub-year long QPOs in jets, which have more cases reported, to LT 
	precession.
\end{abstract}
	

\section{introduction} 

Quasi-periodic oscillations (QPOs) are one type of phenomenon frequently 
observed in X-ray binaries (XRBs) and occasionally in active galactic 
nuclei (AGNs; e.g., \citealt{rm06}; \citealt{im19}). XRB QPOs have been 
extensively 
studied and are believed to reveal the accretion activities of
black holes (BHs; as well as neutron stars). Thus, studies of QPOs can 
provide additional information
for the physical picture and processes of BH accretion, besides
regular studies such as X-ray spectroscopy and multi-band observations.
As the QPO frequency $\nu$ is considered
to scale inversely with BH mass $M_{\rm BH}$, $\nu\sim$10\,Hz 
in XRBs would appear to be several orders of magnitude lower in AGNs
\citep{abr+04};
for example, the first convincing QPO case was $\sim$1\,hr periodicity in 
X-ray emission of the narrow line Seyfert 1 (NLS1) galaxy 
RE J1034+396 \citep{gmw+08,als+14}. It is relatively hard to
detect long periodicities, sub-days to even tens of days expected for AGN
QPOs. Only in recent years, more and more AGN QPO cases
have been reported (see, e.g., \citealt{zw21, tri+24}; references therein), 
which has been enabled by large surveys in time domain at different bands. 
It should be noted that many reported QPOs appeared in AGNs' jet emission,
and thus are radio or $\gamma$-ray cases;
for optical QPOs, there are, for example, those reported in \citet{ote+20}
and \citet{rsz24}.
As jets are coupled to accretion disks, these QPOs in non-thermal emissions
can also potentially reflect accretion activities of AGNs.

\begin{figure*}[!ht]
	\centering
	\includegraphics[width=0.51\textwidth]{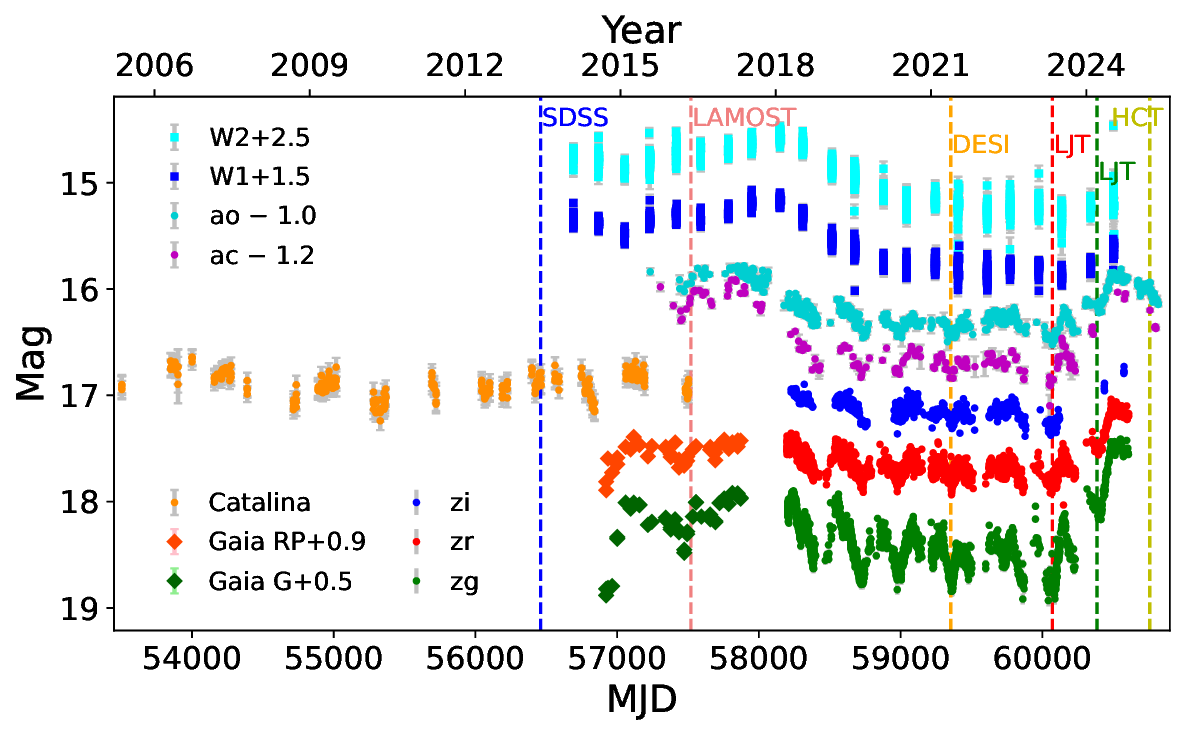}
	\includegraphics[width=0.48\textwidth]{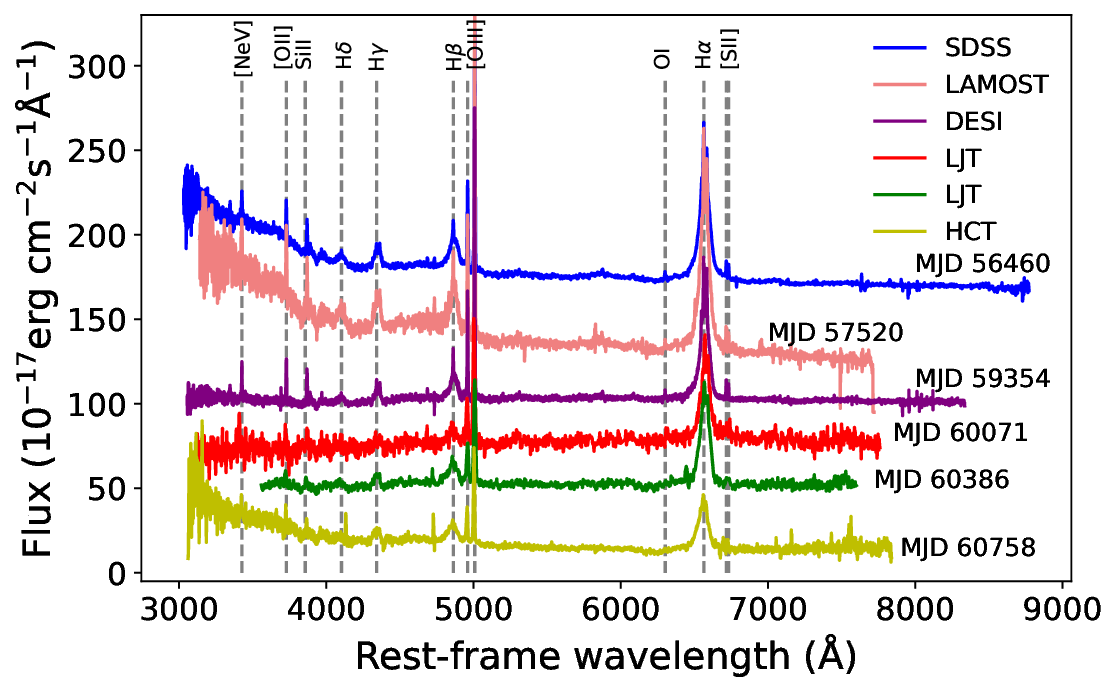}
	\caption{{\it Left:} Long-term optical and MIR LCs of J1626+5120, 
	where the data
	were from CRTS, ZTF, ATLAS, Gaia, and WISE. Vertical dashed lines 
	indicate the epochs of six spectroscopic observations with different 
	telescopes. {\it Right:} six spectra of J1626+5120, which are vertically
	shifted for clarity. The spectrum-taken dates are given. }
	\label{fig:src}
\end{figure*}

QPOs not in obvious association with jets are rare. 
Only several NLS1s were found to exhibit $\sim$1--2\,hr QPOs at X-rays
\citep{gmw+08,als+15,pan+16,gup+18}, besides a 42.2\,d one in type 2 AGN
NGC~4945 \citep{srp20}. At optical, a 44\,d QPO in the NLS1 KIC 9650712
was reported \citep{smb+18}. The hour-long QPOs as well as the 44\,d optical
one follow the relations between $\nu$ and $M_{\rm BH}$ extending from 
high- and low-frequency QPOs (HFQPOs and LFQPOs; see \citealt{im19} for details)
of XRBs respectively
(e.g., \citealt{pan+16,smb+18}). While the origin of the QPOs is under 
discussion (e.g., \citealt{pan+16, rm20}), the scaling relations strongly
suggest similar physical processes occurring in accretion of stellar-mass 
and massive BHs.

In our sample of $\sim$40 AGNs with large optical variations selected from 
Zwicky Transient Facility (ZTF; \citealt{bkg+19}) light-curve (LC) 
data \citep[see][Sect. 1]{zhu+24}, we noted a source,
the Seyfert 1 galaxy J1626+5120 
(R.A. = $16^h26^m11^s.60$, Decl.=$51^{\circ}20^{'}38^{''}.17$, 
J2000.0), showing QPO-like variations. 
We have been obtaining its optical spectra and collecting
archival multi-band data for it. By analyzing the data,
we were able to determine the periodicity $P \simeq 329$\,d and
find evidence for the accretion-disk origin for this QPO.
This rare optical QPO case may provide an explanation for 
the formation mechanims of jet QPOs of similar periodicities. In this paper,
we report our analysis and the results.

Throughout this paper, we adopted cosmological parameters from the Planck 
mission \citep{paa+18}, \emph{$H_0$} = 67.4\,km\,s$^{-1}$ Mpc$^{-1}$ and $\Omega_m$ = 0.315.

\section{Multi-band data and analysis results}
\label{sec:src}

\subsection{Optical light-curve data}

Optical and mid-infrared (MIR) photometric data for J1626+5120 from 
several archives were collected, which included ZTF $g$ ($zg$),
$r$ ($zr$), and $i$ ($zi$) bands, Asteroid Terrestrial-impact Last Alert System 
(ATLAS; \citealt{tdh+18}) $c$ ($ac$) and $o$ ($ao$) bands,
Catalina Real-time Transient Survey (CRTS; \citealt{ddm+09}) 
$V$-band, Gaia Data Release 3 (DR3; \citealt{gaia16,gaiaDR3})
in $G$ and $RP$ bands,
and NEOWISE W1 ($3.4\,\mu$m) and W2 ($4.6\,\mu$m) bands
\citep{Mainzer+11,neowise}. 
The multi-band LCs (left panel of Figure~\ref{fig:src}) resulting from the data
show several-year long variations in the low-cadence data (MIR W1 and W2, 
and probably CRTS $V$ as well),
but in the high-cadence ZTF and ATLAS data, especially $zg$, year-long 
variations with amplitudes as large as $\sim$1\,mag are clearly visible.

\subsection{Optical spectrum data}
\label{subsec:spec}

We obtained six optical spectra between MJD 56460 and 60758
(the right panel of Figure~\ref{fig:src}).
Three were respectively from the Sloan Digital Sky Survey 
(SDSS; MJD 56460; \citealt{aaa+09}), the Large Sky Area Multi-Object Fiber 
Spectroscopic Telescope (LAMOST; MJD 57520; \citealt{lamost12}), and the Dark 
Energy Spectroscopic Instrument (DESI; MJD 59354; \citealt{desi16}). 
The other three were taken by us, with two with the Lijiang 2.4-m Telescope 
(LJT; \citealt{wbf+19}) on MJD 60071 and 60386 (40\,min and 50\,min
exposures respectively)
and one with the 2-m Himalayan Chandra Telescope (HCT; \citealt{csp02}) on 
MJD 60758 (two 1-hr exposures).
The Yunnan Faint Object Spectrograph and Camera (YFOSC) with Grism 3 and 
a $2\farcs5$ slit was used for obtaining the LJT spectra, and
the Hanle Faint Object Spectrograph and Camera (HFOSC) with two settings,
Grism 7 with a $0\farcs77$ slit and Grism 8 with a $1\farcs15$ slit
(in order to have a wide wavelength coverage), was
used for obtaining the HCT spectrum.  

\begin{figure}[!ht]
	\centering
	\includegraphics[width=0.49\textwidth]{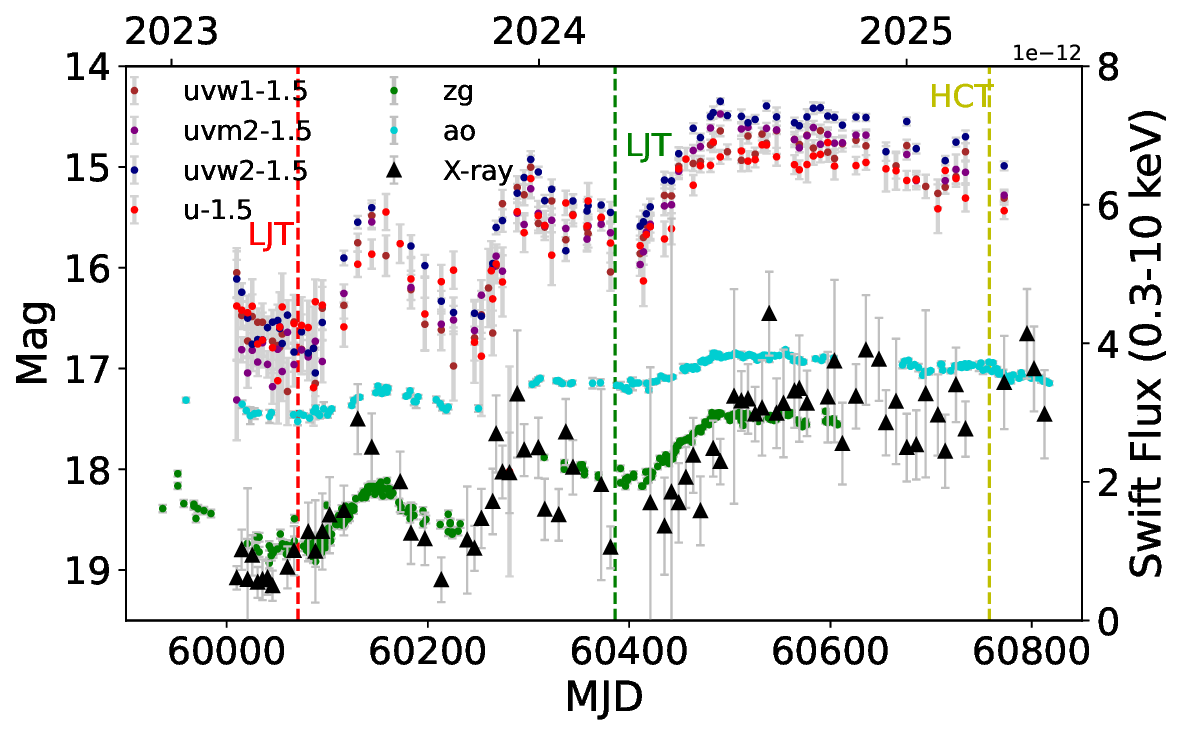}
	\caption{Swift XRT and UVOT LCs of J1626+5120, 
	for which $zg$ and $ao$ LCs are plotted for comparison. 
	Vertical dashed lines mark epochs of LJT (red and green) and HCT 
	(gold) spectroscopic observations.
	}
	\label{fig:lc_swift}
\end{figure}

Based on the SDSS spectrum,
J1626+5120 was classified as a Seyfert 1 galaxy
at redshift $z=0.1785$ \citep{sdssdr12}.
No significant line profile changes were detected 
(see Figure~\ref{fig:spec_hab}). 
The spectra indicate that the LC variations were due to the continuum 
at the wavelength of $\lesssim$4000\,\AA, the region showing
significantly different fluxes between the bright-state spectra 
(SDSS, LAMOST, HCT) and dim-state 
spectra (DESI, LJT). It is interesting to note that the DESI and LJT
spectra were taken near the bottom of the year-long variations 
(Figure~\ref{fig:src}).

As the DESI spectrum has the finest pixel scale 
(0.67\,\AA\,pixel$^{-1}$) over a wide wavelength range (3055--8336\,\AA), 
better than that of SDSS (1.25\,\AA\,pixel$^{-1}$), 
LAMOST (1.17\,\AA\,pixel$^{-1}$), LJT ($\sim$2.45\,\AA\,pixel$^{-1}$), and 
HCT (1.23\,\AA\,pixel$^{-1}$), while having a sufficient 
signal-to-noise ratio with an average value of $\sim$15.6,
we fit it using \texttt{PyQSOFit} \citep{gsw18} 
and found that the optical emission was dominated by the AGN with no 
significant contribution from the host galaxy (Figure~\ref{fig:desi}).
From the fitting results, 
the full-width at half maximum (FWHM) of the H$\beta$ line
and the continuum luminosity at 5100\,\AA\ ($L_{\lambda}(5100\,\text{\AA})$),
we estimated $M_{\mathrm{BH}}$
from the single-epoch virial mass relation given by \citet{vp06},
$\log_{10}(M_{\mathrm{BH}}/M_{\odot}) \simeq 8.01^{+0.10}_{-0.11}$.
The Eddington ratio 
$\dot{m}_{\text{Edd}} = L_{\text{bol}}/L_{\text{Edd}} \simeq 0.043\pm0.012$ was 
also obtained, where the bolometric luminosity $L_{\text{bol}}$ was estimated 
from $L_{\lambda}(5100\,\text{\AA})$ 
using a bolometric correction of 9.26 
(\citealt{rls+06}; applicable for $z<0.8$) and the Eddington luminosity
$L_{\text{Edd}} = 1.38 \times 10^{38} (M_{\mathrm{BH}}/M_{\odot})$ erg s$^{-1}$.

\subsection{{\it Swift} X-ray and ultraviolet data}

There were 107 observations of J1626+5120 conducted with 
{\it the Neil Gehrels Swift Observatory} ({\it Swift}; \citealt{swift04})
between March 7, 2023 and May 17, 2025. The observations were
in 0.3--10\,keV with the X-Ray Telescope (XRT; \citealt{xrt05})
and at Ultraviolet (UV) bands with the Ultraviolet Optical Telescope 
(UVOT; \citealt{rkm+05, poo+08}). 
Most of the XRT and UVOT exposures range from 500\,s to 3000\,s.

The XRT Photon Counting (PC) mode data were processed through the UK {\it Swift}
Science Data Centre (UKSSDC) automated pipeline \citep{ebp+09}. 
An absorbed power-law (PL) model was used to fit the spectra of the source,
where the hydrogen column density $N_H$ was fixed at the Galactic value
$1.86 \times 10^{20}$\, cm$^{-2}$ \citep{kbh+05}. The source's 
emission showed a weak softer-when-brighter trend
(see Figure~\ref{fig:xindex}). Given the quality of the data, 
no soft component, in addition to the PL, could be clearly determined.
Considering the uncertainties of the flux measurements 
(cf., Figure~\ref{fig:xindex}) and the distribution of the XRT exposure times, 
we only used those data with exposure times $>$1400\,s in our following 
analysis.
The selection resulted in 76 data points.


The UV data used in our analysis were 95 exposures at filters 
$uvw2$ (central wavelength 
$\lambda_c=1928\,$\AA, FWHM$=657\,$\AA), $uvm2$ ($\lambda_c=2246\,$\AA, 
FWHM$=498\,$\AA), $uvw1$ ($\lambda_c=2600\,$\AA, FWHM$=693\,$\AA),
or $u$ ($\lambda_c=3465$\,\AA, FWHM$=785$\,\AA).
The data were processed using the \texttt{uvotsource} tool. For each image, 
a $5^{\prime\prime}$ radius circular region centered at the source was used 
for extracting the source's photons, and the background was measured from a 
nearby $25^{\prime\prime}$ radius circular region.

The {\it Swift} measurements are shown in the left panel of
Figure~\ref{fig:lc_swift}, along
with optical $zg$ and $ao$ LCs for comparison. As can be seen, 
the X-ray and UV fluxes showed correlated variations.

\begin{figure*}[!ht]
	\centering
	\includegraphics[width=0.45\textwidth]{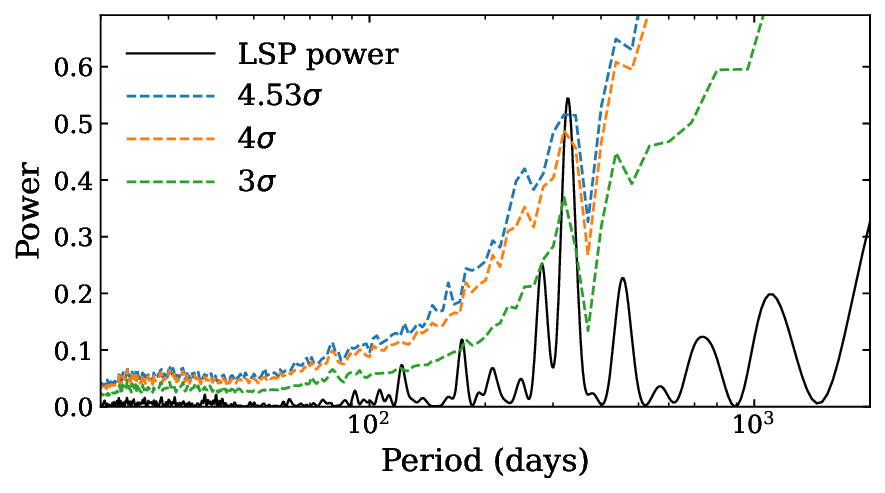}
	\includegraphics[width=0.48\textwidth]{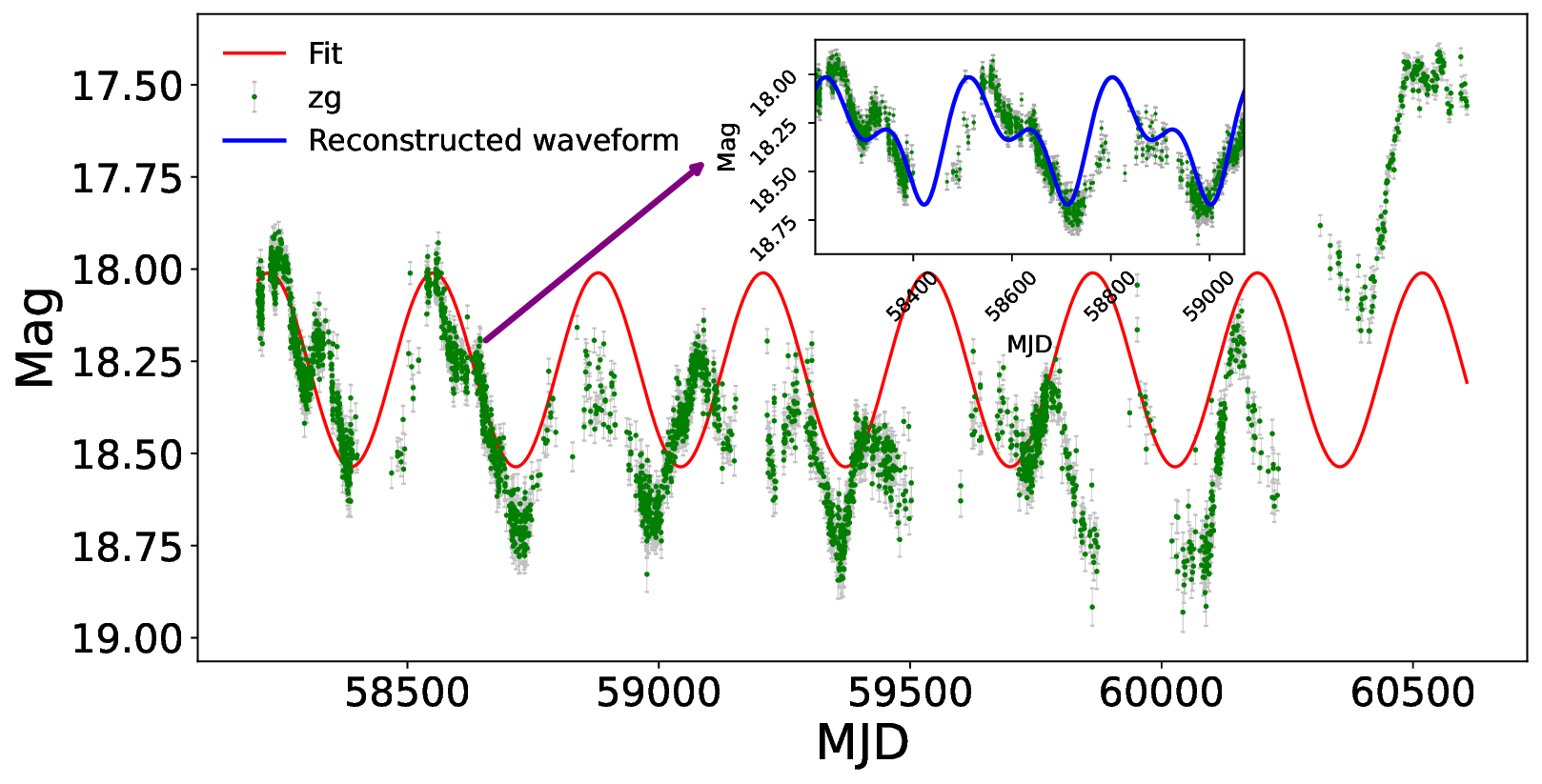}
	\caption
	{{\it Left:} LSP power spectrum (black solid) from the $zg$ LC of J126+5120. 
	The dashed lines indicate the $3\sigma$, $4\sigma$, and $4.53\sigma$ 
	significance levels. {\it Right:} $zg$ LC (green data points) of 
	J1626+5120 and a sinusoidal 
	fit (red solid line) to the LC, with the latter used to help 
	illustrate the QPO signal.
	At $P = 329$\,d, the sinusoid has a semi-amplitude of 0.26\,mag. 
	The inset shows the first three cycles of the QPO variations, 
	which have a waveform 
	(described with the blue solid line; \citealt{iv15}) similar 
	to that seen in XRBs.
	}
	\label{fig:lspp}
\end{figure*}

\section{Temporal variability analysis}


\subsection{Periodicity determination}\label{subsec:pd}

Among the optical LCs, the $zg$ one shows clear modulation 
(Figure~\ref{fig:src}). By performing timing analysis to the LCs at
different bands (e.g., $zg$, $zr$, $zi$, $ao$), 
a QPO signal was found to be most significant in the $zg$ data. 
We therefore focused on the $zg$ band. The method 
used to determine the signal was the Lomb-Scargle Periodogram 
(LSP; \citealt{lomb76,scargle82}). The analysis with the method resulted in
a significant QPO at $\simeq329\pm 18$\,d (Figure~\ref{fig:lspp}),  
which was estimated from a Gaussian fit to the power peak in the LSP spectrum 
with the uncertainty being the FWHM of the fit. 
To visually illustrate the QPO modulation, 
a sinusoidal fit, with 329\,d period, to the $zg$ LC is shown in the right panel of
Figure~\ref{fig:lspp}.
It can be noted that the overall phase-averaged brightness had small 
decreases/variations and then jumped by $\sim -$0.7\,mag in the end of the LC.

To evaluate the statistical significance of this detection against red-noise 
fluctuations, we generated $3 \times 10^5$ synthetic LCs replicating 
the statistical properties (i.e., the power spectral density and the flux probability distribution function)
of the observed $zg$ data, 
following \citet{emp13}. Applying the LSP analysis to each synthetic LC,
we determined significance thresholds from the resulting distribution 
of maximum noise power levels. Specifically, the $3\sigma$,
$4\sigma$, and $4.53\sigma$ significance 
curves are overlaid on the LSP power spectrum in the left panel of Figure~\ref{fig:lspp},
and the power of the QPO reaches the $4.53\sigma$ curve.

As a cross-check, we also employed the Weighted Wavelet Z-transform 
(WWZ; \citealt{wwz96}), where the Python code was from \citet{wwzcode},
to investigate the temporal evolution of the QPO signal. The result, 
presented in Appendix Figure~\ref{fig:wwz}, reveals that the $\sim$329\,d
signal was persistent throughout the entire time span of the $zg$ data.

\subsection{Cross-correlation analysis}\label{sec:corr_analy}

\begin{figure}[!ht]
	\centering
	\includegraphics[width=0.48\textwidth]{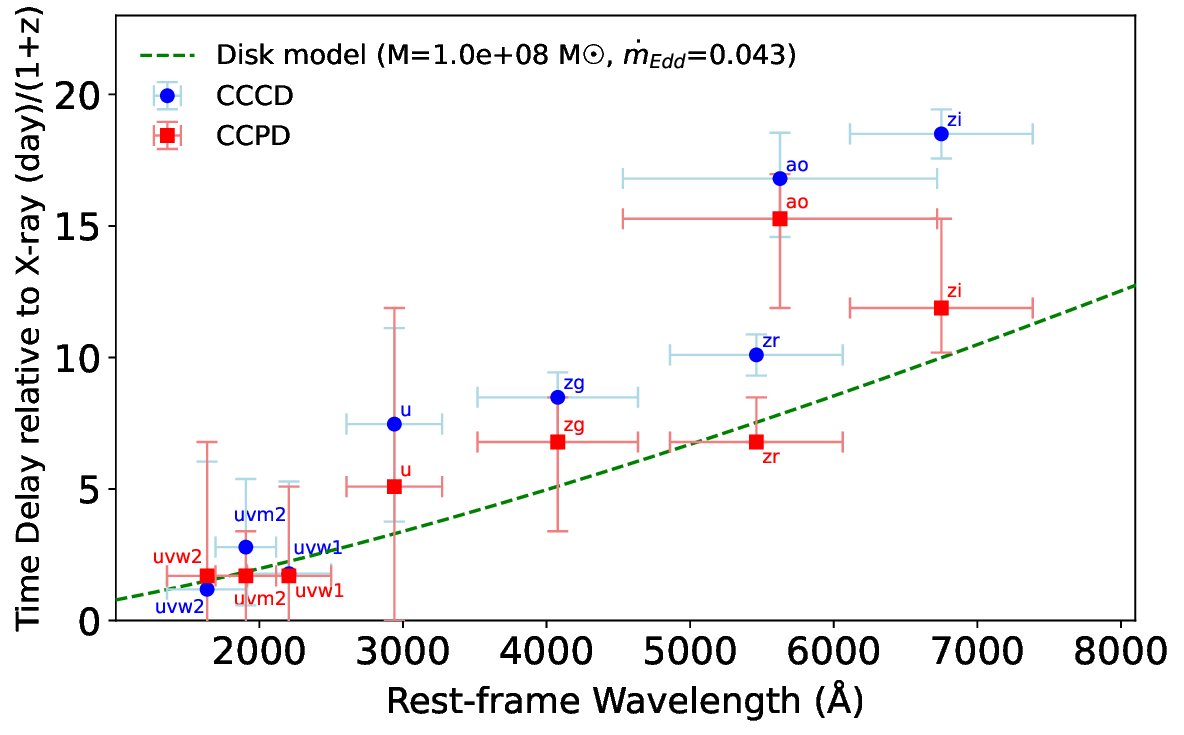}
	\caption{Time lags of the UV
		and optical LCs with respect to the X-ray LC determined from 
		cross-correlation analysis. A green dashed curve 
		indicates the time delay expected from a disk continuum reprocessing
		model.
	}
	\label{fig:lag}
\end{figure}

To examine the correlations between LCs of the different bands, we
used the \texttt{PyCCF} Python package \citep{pyccf18} for the analysis.
The XRT LC was set as the reference with zero time lag. The time lags of
the UV and optical LCs were determined from the Cross-Correlation Peak 
Distribution (CCPD) and Centroid Distribution (CCCD) provided in {\tt PyCCF}. 
Our initial analysis indicated that direct cross-correlations of the UV
or optical LCs with the X-ray LC resulted in large uncertainties,
and we thus adopted the following strategy:
1) the $uvw2$ was measured directly against the X-ray LC; 
2) the lags for the $uvm2$, $uvw1$, $u$, and $zg$ bands were first 
measured relative to the $uvw2$, and the final lags relative to the X-ray
were obtained by adding the $uvw2$ lag;
3) similarly, the lags for the $zr$, $zi$, and $ao$
bands were first measured relative to the $zg$, and
their final lags relative to the X-ray
were calculated by adding the $zg$ lag.
Time lag uncertainties were estimated using the Flux 
Randomization/Random Subsample Selection (FR/RSS) Monte Carlo method. This 
technique creates thousands of LC realizations by modifying fluxes within 
their errors and selecting random data subsets, and the final lag and 
1$\sigma$ uncertainty are derived from the resulting CCCD and CCPD 
distributions.

Individual analysis plots are provided in Figure~\ref{fig:multi-lag}. 
The obtained time lagsare summarized in Table~\ref{tab:lags} and plotted as a function of effective 
wavelength in Figure~\ref{fig:lag}. 
A trend of larger positive lags at longer wavelengths is seen. 
Such a trend is often explained with a continuum-reprocessing disk model.
We calculated the model lags, $\tau(\lambda)$, based on Equation~(12) in
\citet{fdb+16}. For J1626+5120, we adopted $M_{\text{BH}} = 10^8\ M_{\odot}$ 
and $\dot{m}_{\text{Edd}} = 0.043$ (Section~\ref{subsec:spec}), 
and assumed radiative efficiency $\eta = 0.1$, external to internal heating
ratio $\kappa = 1$, and scaling factor $X = 5.04 $ \citep{tk18}. 
This model has a lag-wavelength relationship of 
$\tau(\lambda) \propto \lambda^{4/3}$. 
As shown in Figure~\ref{fig:lag}, the observed lags are slightly larger 
than the model ones. The discrepancy, which suggests larger disk sizes 
than the disk model, is commonly reported in AGN reverberation 
mapping studies (e.g., \citealt{fdb+16,gbw+22}). On the other hand,
recent theoretical studies have shown that more sophisticated X-ray 
reverberation models or disk models can account for large lags through 
a more physical treatment of the reprocessing mechanism 
(e.g., \citealt{ppp+24,phh+25}).

\section{Discussion}
\label{sec:dis}

The type 1 AGN J1626+5120 has shown intriguing LC variations, and our analysis
has indicated that in the well-sampled $zg$ LC, 
a $P \simeq 329$\,d QPO existed at a $\sim 4.53\sigma$ significance.
The six optical spectra taken over the course of the long-term LCs have revealed
that the flux changes were mainly caused by the varied blue continuum, presumably
arising from the accretion disk.
Also, the {\it Swift} X-ray and UV flux measurements,
although covering $\sim$2\,yr of the last portion of the long-term LCs,
showed variations correlated with the optical ones, and the time lags can
be approximately described with a reprocessing disk model. Combining these
properties together, we believe that this is likely a QPO case
related to physical processes at the inner region of the accretion disk 
in J1626+5120.

\begin{table}
	\centering
	\caption{Time lags relative to X-ray emission}
	\label{tab:lags}
\begin{tabular}{lcccc}
\hline
Band & Wavelength & FWHM & CCCD Lag & CCPD Lag \\
 & (\AA) & (\AA) & (day)   & (day)   \\
\hline
$uvw2$ & 1928 & 657 & $1.40^{+5.72}_{-5.57}$  & $2.0^{+6.0}_{-12.0}$ \\
$uvw1$ & 2600 & 693 & $2.10^{+4.13}_{-3.69}$  & $2.0^{+4.0}_{-4.0}$  \\
$uvm2$ & 2246 & 698 & $3.29^{+3.05}_{-2.61}$  & $2.0^{+2.0}_{-4.0}$  \\
$u$    & 3465 & 785 & $8.80^{+4.30}_{-4.37}$  & $6.0^{+8.0}_{-6.0}$  \\
$zg$   & 4806 & 1317& $10.00^{+1.12}_{-2.01}$ & $8.0^{+2.0}_{-4.0}$  \\
$zr$   & 6436 & 1418& $11.90^{+0.92}_{-0.93}$ & $8.0^{+2.0}_{-0.0}$  \\
$zi$   & 7954 & 1500& $21.80^{+1.10}_{-1.10}$ & $14.0^{+4.0}_{-2.0}$ \\
$ao$   & 6630 & 2578& $19.80^{+2.05}_{-2.62}$ & $18.0^{+2.0}_{-4.0}$ \\
		\hline
	\end{tabular}
\end{table}	

Considering $M_{\rm BH}\simeq 10^8\ M_{\sun}$, the gravitational radius
of the BH $R_g = GM_{\rm BH}/c^2 \simeq 1.5\times 10^{13}$\,cm. The Keplerian
period of the innermost stable circular orbit (ISCO)
is $\simeq$13\,hr, which is too short to be related to J1626+5120's QPO
for discussion of the QPO origin
(e.g., \citealt{gmw+08,tm21}). A truncated disk possibility
(\citealt{cr12,lrj+13}) has been discussed
(see, e.g., \citealt{tm21}), based on which the disk's inner edge would be
at $\sim 380R_g$ in order to match the 329\,d period. Such a large truncation 
is unusual and would be hard to explain the varied blue continuum (which 
is likely the cause of the modulation). 
The period of an LFQPO, if scaled with the relationship,
$\nu \propto 51.9$\,Hz $(M_{\rm BH}/M_{\sun})^{-1}$ given in \citet{smb+18},
would be $\sim$22\,d, also being too short. However,
the LFQPOs of XRBs are observed to have a wide range of frequencies, from
a few mHz to $\sim$30\,Hz (see details in \citealt{im19}). If this is also true 
to AGN QPOs, this QPO could simply correspond to a sub-Hz one in XRBs.
In addition, we noted that the first three cycles of the QPO variations show
a secondary peak (see in the right panel of Figure~\ref{fig:lspp}). This shape of a waveform was seen
in QPOs of XRBs and is the result of the addition of two harmonics \citep{iv15}. 
Given these, we are inclined to identify the QPO as a counterpart of LFQPOs in XRBs.

For LFQPOs in XRBs and $\gamma$-ray bursts as well,
the Lense-Thirring (LT) precession of the inner accretion flow is a compelling 
scenario widely discussed \citep{sv98, idf09, im19, tm21}.
In the scenario,
the frame-dragging effect caused by a spinning BH induces
a geometrically thick inner flow or a warped inner region of a thin disk 
to precess, resulting in modulated observed emission.
The timescale of LT precession ($P_{\text{LT}}$) for a ring at a radius 
$R_p = r_p R_g$ (where $r_p$ is the dimensionless radius in units
of $R_g$) 
around a Kerr BH with dimensionless spin parameter $a_*$ can 
be estimated as \citep{wil72,sv98} 
$P_{\text{LT}} \approx 0.018$\,d $(M_{\text{BH}}/10^8\ M_{\odot}) (r_p^3/a_*)$.
For J1626+5120's QPO, 
$r_p^3/a_* \approx 15510$ ($P_{\text{LT}} = P/(1+z) \simeq 279$\,d) is 
obtained. 
Considering $0.014 < a_* < 1$, $r_p$ ranges from 6, radius of the ISCO,
to 24.  The radius range are reasonable values, similar to that estimated from
the QPOs in tidal disruption events (e.g., \citealt{pas+24}).
In addition, $a_*$ can be estimated through the accretion efficiency 
method (e.g., \citealt{ztb+20}), as the radiative efficiency $\eta$ is related 
to $a_*$ from $\eta(a_*) = 1 - E_{\rm ISCO}(a_*)$, where $E_{\rm ISCO}$ is 
the specific energy at ISCO. For a standard thin disk, $\eta$ ranges from 
$\sim 0.057$ ($a_*=0$) to $\sim 0.32$ ($a_*\approx 1$) \citep{bpt72}. 
For J1626+5120, its $\eta \approx 0.09$, given from 
the relation $\eta \approx 0.089 (M_{\text{BH}}/10^8 M_{\odot})^{0.52}$ 
\citep{dl+11},
and this $\eta$ value corresponds to a moderately high spin of 
$a_* \approx 0.6$ (also see Fig. 6 in \citealt{pbn25}). Thus,
$r_p \approx 21\,R_g$ in J1626+5120.

This J1626+5120 QPO potentially could be a key case for helping understand 
those hundreds-day long jet QPOs, such as $\sim$120/150\,d in blazar J1359+4011
\citep{khm+13}, 176\,d in NLS1 galaxy J0849+5108 \citep{zw21},
or even 345/386\,d in NGC~1275 \citep{zha+22}. The sub-year periods are too
short for another often-discussed scenario, signals of binary supermassive BH 
(SMBH) systems.  Taking J1626+5120 as an example, if it contains two SMBHs
with equal mass, the separation would be $\simeq$0.002\,pc, already having 
passed the $\sim$Gyr-long hardening phase (e.g., \citealt{yu02,col14});
for comparison, several candidate binary SMBH systems have several-year long
periodicities (e.g., \citealt{val+08,gra+15}).
It has been shown
from numerical simulations that jets, along with their disks as a system, 
can undergo LT precession \citep{lht+18}. 
We note that the BHs in the radio-loud AGNs listed above
have similar $M_{\rm BH}$ ($\sim 10^{7-8}\ M_{\sun}$).
Thus due to LT precession, the sources would have similar QPO periods,
while appearing in their jets.  Detailed property studies of AGNs 
with sub-year QPOs may help fully examine this possibility.

\begin{acknowledgments}
This work was based on observations obtained with the Samuel Oschin Telescope 
48-inch and the 60-inch Telescope at the Palomar Observatory as part of the 
Zwicky Transient Facility project. ZTF is supported by the National Science
Foundation under Grant No. AST-2034437 and a collaboration including Caltech, 
IPAC, the Weizmann Institute for Science, the Oskar Klein Center at Stockholm 
University, the University of Maryland, Deutsches Elektronen-Synchrotron
and Humboldt University, the TANGO Consortium of Taiwan, the University of 
Wisconsin at Milwaukee, Trinity College Dublin, Lawrence Livermore National 
Laboratories, and IN2P3, France. Operations are conducted by COO, IPAC, and UW.

This work has made use of data from the European Space Agency (ESA) mission
{\it Gaia} (\url{https://www.cosmos.esa.int/gaia}), processed by the {\it Gaia}
Data Processing and Analysis Consortium (DPAC,
\url{https://www.cosmos.esa.int/web/gaia/dpac/consortium}). Funding for the DPAC
has been provided by national institutions, in particular the institutions
participating in the {\it Gaia} Multilateral Agreement. We acknowledge ESA {\it Gaia}, DPAC and the Photometric Science Alerts Team (\url{http://gsaweb.ast.cam.ac.uk/alerts}).

This publication makes use of data products from the Wide-field Infrared Survey Explorer, which is a joint project of the University of California, Los Angeles, and the Jet Propulsion Laboratory/California Institute of Technology, funded by the National Aeronautics and Space Administration. This publication also makes use of data products from NEOWISE, which is a project of the Jet Propulsion Laboratory/California Institute of Technology, funded by the Planetary Science Division of the National Aeronautics and Space Administration.

This study makes use of data obtained from the 2-m Himalayan Chandra Telescope 
(HCT). We thank the staff at IAO, Hanle, and CREST, Hosakote, operated by the 
Indian Institute of Astrophysics, Bengaluru (India), for their support in 
	facilitating these observations.

	We thank the anonymous referee for very detailed and helpful
	comments and Y. Luo  for helping with theoretical understanding.
This research is supported by the National Natural Science Foundation of 
	China (12273033).  L.Z. acknowledges the support of the science 
	research program for graduate students of Yunnan University (KC-24249083).
\end{acknowledgments}

\appendix

\restartappendixnumbering

\section{Supplementary Figures}

\begin{figure}[!ht]
	\centering
	\includegraphics[width=0.5\textwidth]{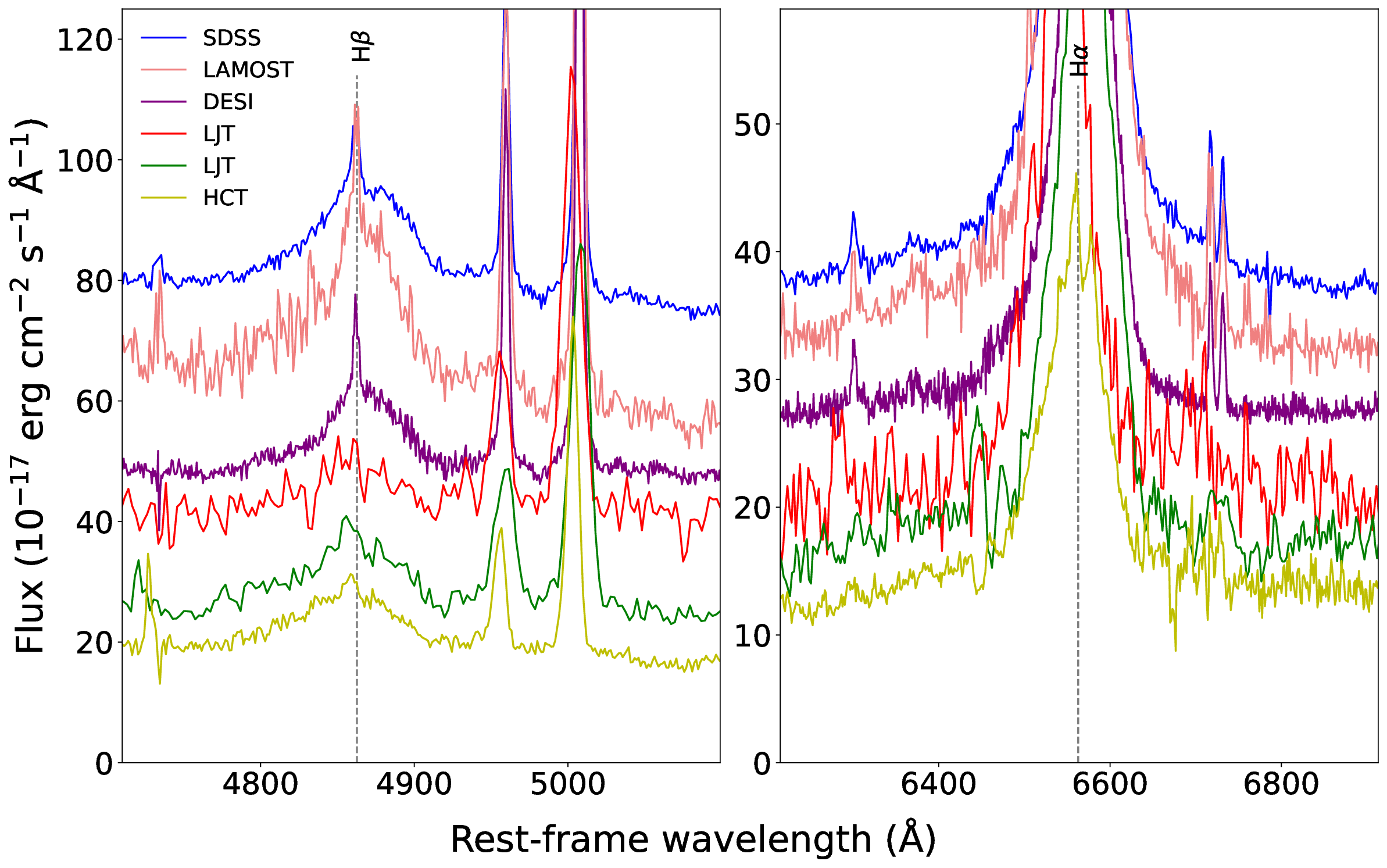}
	\caption
	{Zoomed-in views of the optical spectra of J1626+5120. 
	\textit{Left:} H$\beta$ and [O III] region. \textit{Right:} 
	H$\alpha$ region. 
	The spectra are vertically shifted for clarity.  }
	\label{fig:spec_hab}
\end{figure}

\begin{figure*}
	\centering
	\includegraphics[width=0.94\linewidth]{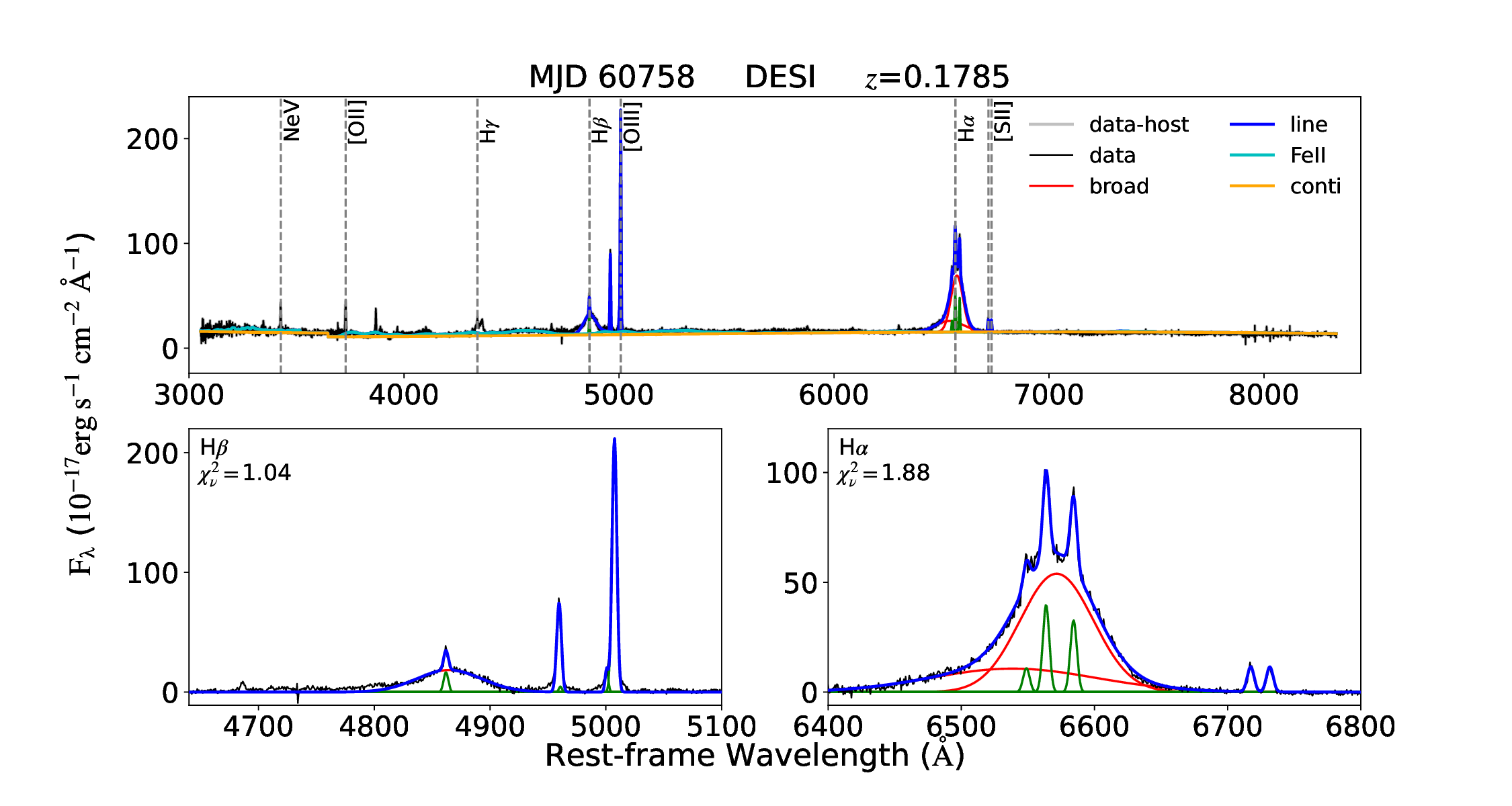}
	\caption{Fit to the DESI spectrum of J1626+5120 with PyQSOFit. 
	FWHM(H$\beta$)$ = 4046\pm44$\,km\,s$^{-1}$ and 
	$L_{\lambda}$(5100\,\AA) $= 10^{43.8\pm4.3}$\,erg\,s$^{-1}$ were
	obtained from the fitting.}
	\label{fig:desi}
\end{figure*}

The spectra of the H$\beta$ and [O III] region and H$\alpha$ region 
are shown
in Figure~\ref{fig:spec_hab}, and detailes of the fitting to the DESI spectrum
with PyQSOFit are shown in Figure~\ref{fig:desi}.

\begin{figure}[!ht]
	\centering
	\includegraphics[width=0.5\textwidth]{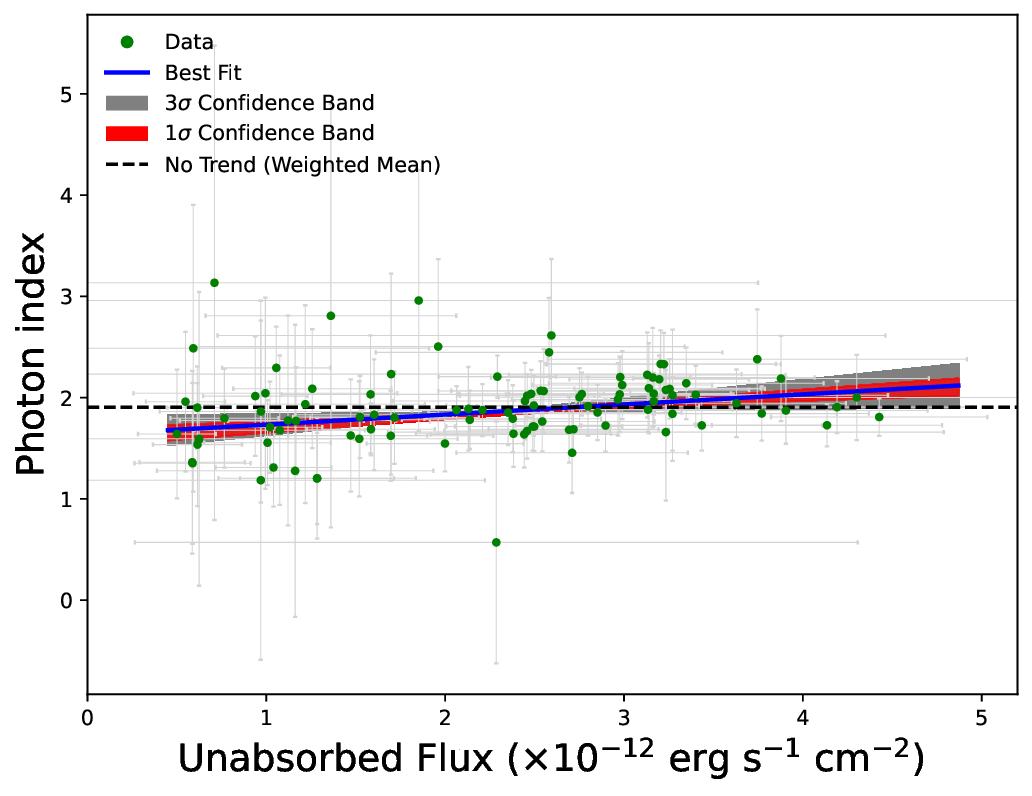}
	\caption
	{Photon index ($\Gamma$) versus unabsorbed 0.3--10 keV X-ray 
	flux ($F_X^u$) 
	for J1626+5120 obtained from {\it Swift} XRT observations. 
	The blue solid line represents the best linear fit, which 
	shows a weak softer-when-brighter trend (with a 3.3$\sigma$ 
	significance): $F_X^u$/$10^{-12}$ $\sim$ (0.10$\pm$0.03) $\times \Gamma + (1.63\pm0.08)$.  
	The black dashed line marks the weighted mean of
	the $\Gamma$ values (with their uncertainties as the weights), which is 
	$\simeq$1.91.}
	\label{fig:xindex}
\end{figure}

In Figure~\ref{fig:xindex}, photon index versus unabsorbed 0.3--10\,keV X-ray
flux, obtained from 76 XRT observations with exposures $>$1400\,s, is shown. 
A weak softer-when-brighter trend was derived by performing a weighted 
orthogonal regression which accounts for uncertainties in both variables.

\begin{figure}[!ht]
	\centering
	\includegraphics[width=0.5\textwidth]{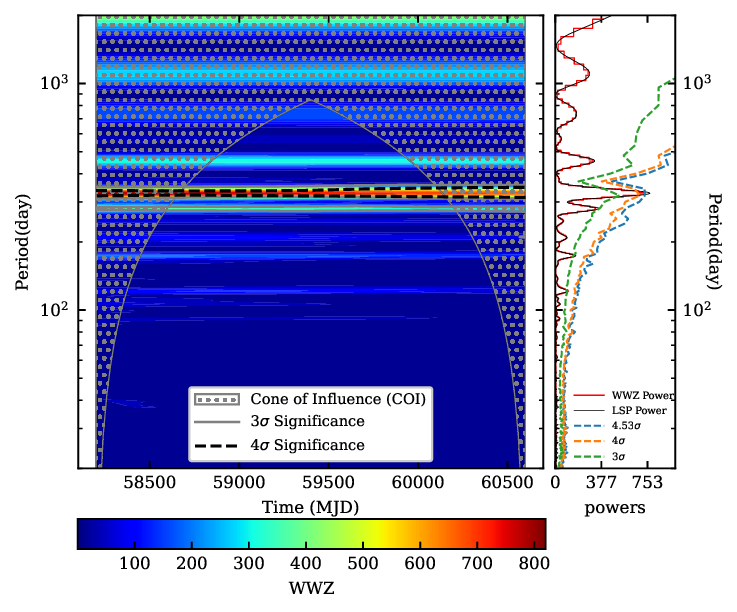}
	\caption
	{Periodicity analysis of the $zg$ LC of J1626+5120. \textit{Left:} 
	color-scaled WWZ power spectrum, with the $3\sigma$ and 
	$4\sigma$ confidence contours (gray and black dashed lines, 
	respectively) overlaid. The gray dotted region indicates the Cone of 
	Influence (COI; \citealt{tc98, tzm+20}), within which the results are 
	affected by edge effects. The main QPO signal at $\sim$329\,d is 
	located outside the COI, confirming the detection is reliable. 
	\textit{Right:} Same as the left panel of Figure~\ref{fig:lspp}, 
	with the time-averaged WWZ power spectrum (red histogram) plotted.
	}
	\label{fig:wwz}
\end{figure}

In Figure~\ref{fig:wwz}, the WWZ periodicity analysis results for J1626+5120
are shown. Besides the 329\,d signal, we noted that there was an additional
possible signal at $\sim$280\,d, which reached 3$\sigma$. Detailed exmination
of the $zg$ LC
suggests that this signal was likely caused by the first three cycles of the
LC modulation (cf., the right panel of Figure~\ref{fig:lspp}).

\begin{figure}[!ht]
	\centering
	\setlength{\tabcolsep}{5pt}
	\begin{tabular}{cc}
		\includegraphics[width=0.45\textwidth]{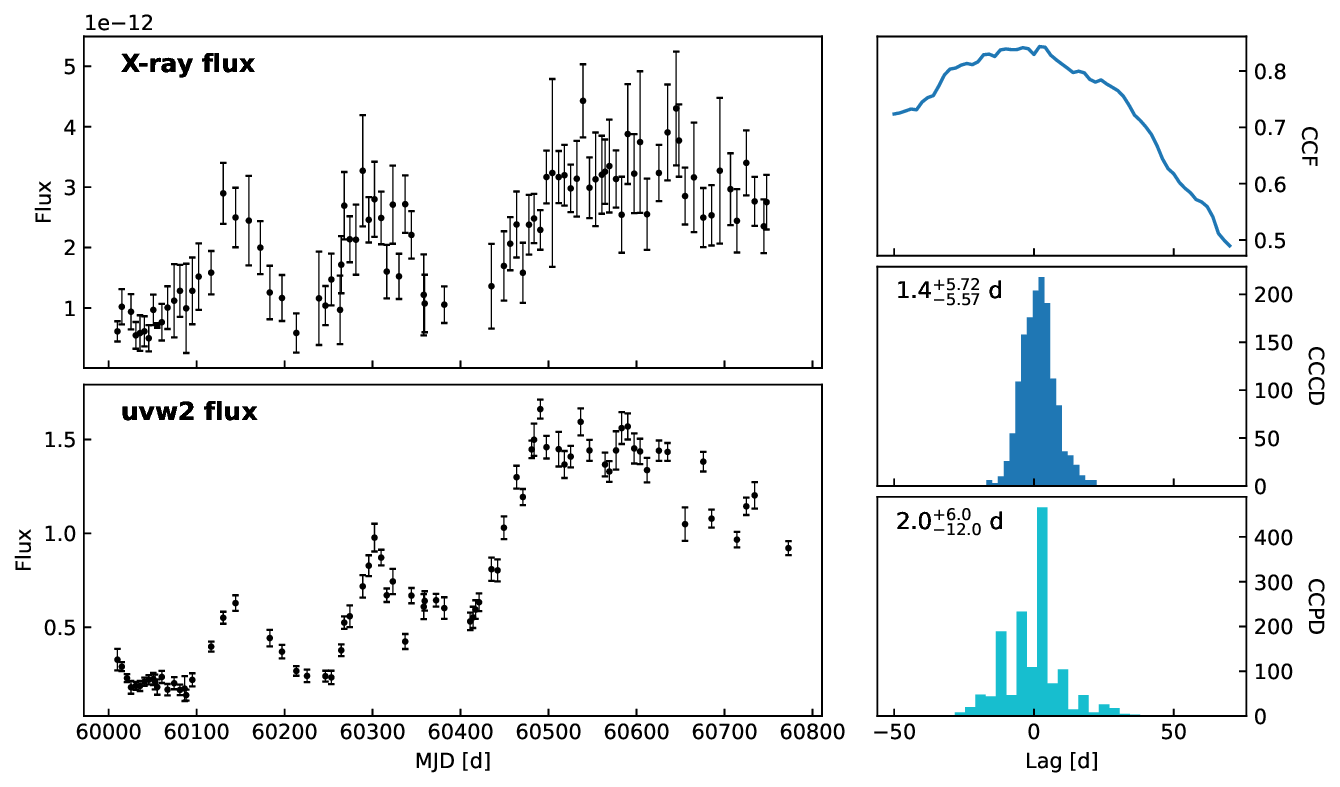} &
		\includegraphics[width=0.45\textwidth]{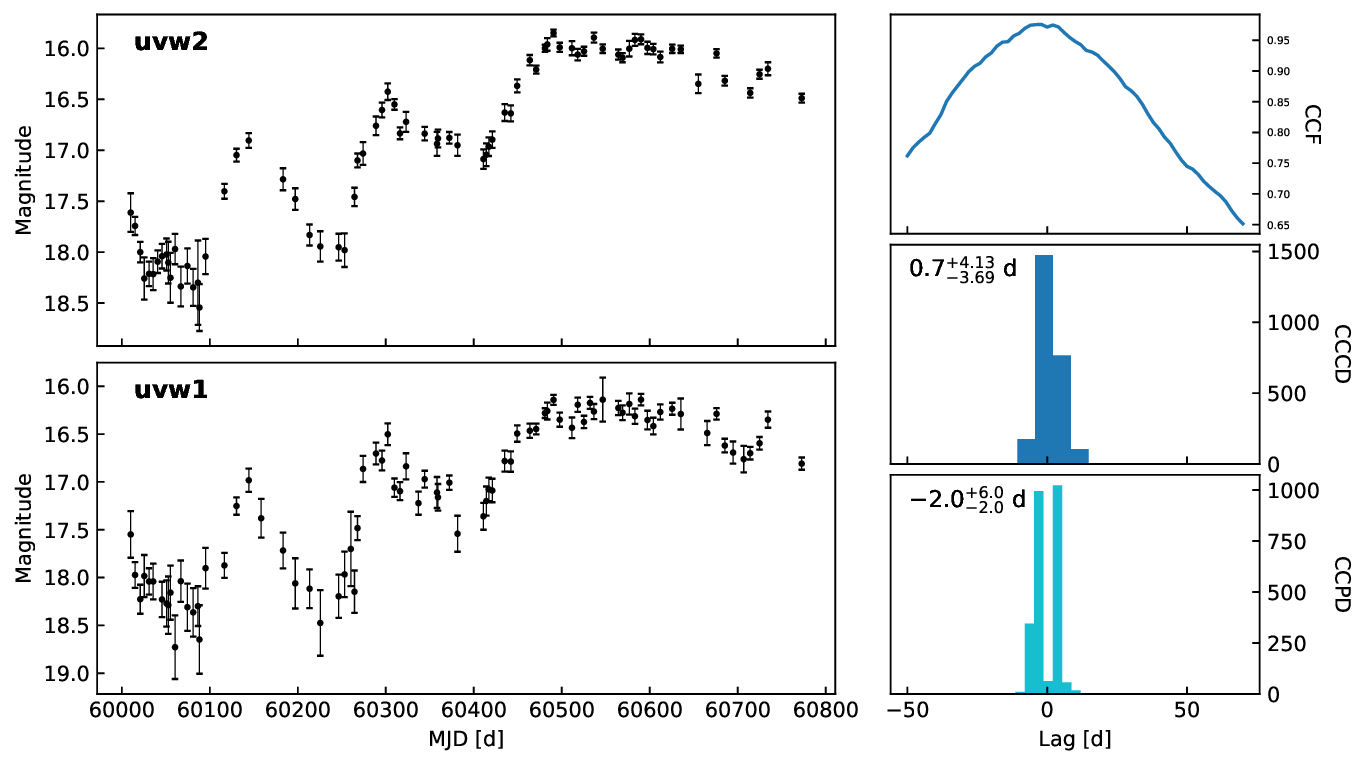}    \\[2pt]
		\includegraphics[width=0.45\textwidth]{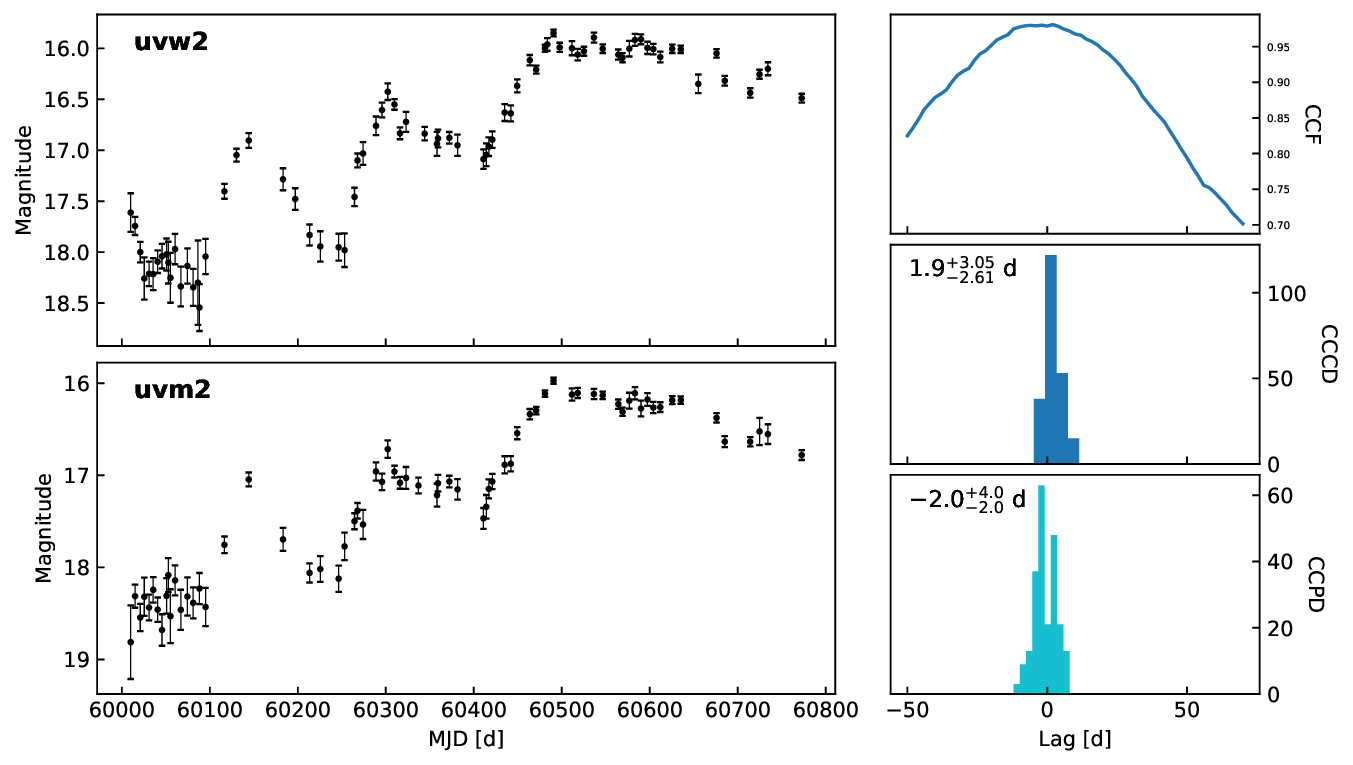}    & 
		\includegraphics[width=0.45\textwidth]{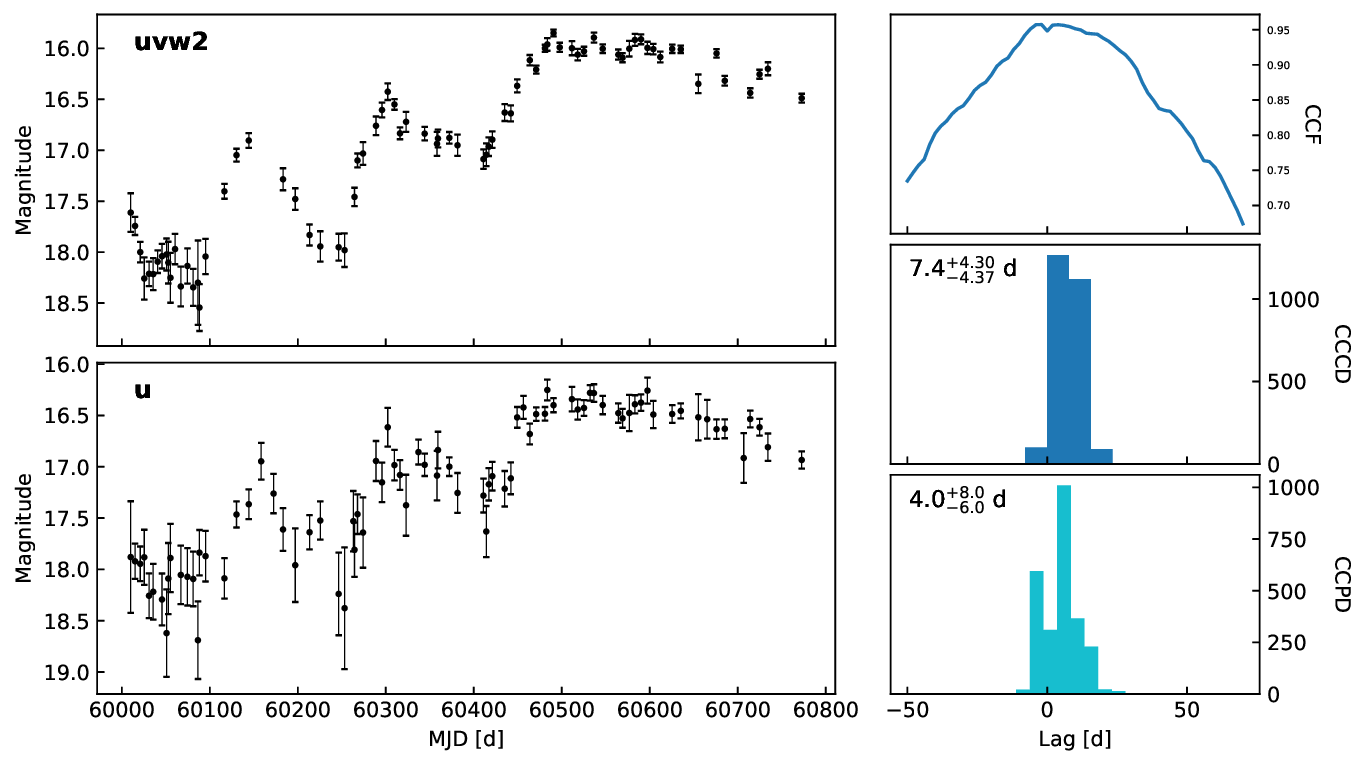}    \\[2pt]  
		\includegraphics[width=0.45\textwidth]{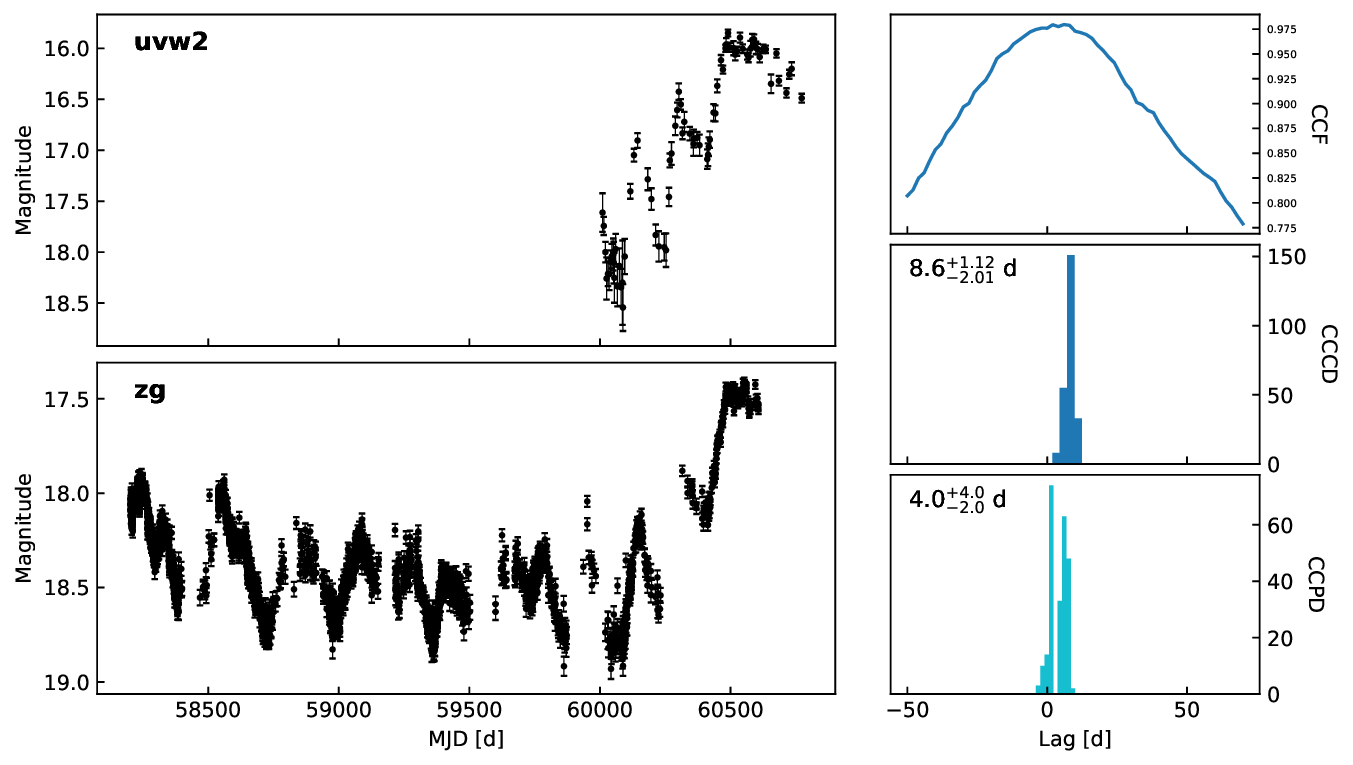}     &
		\includegraphics[width=0.45\textwidth]{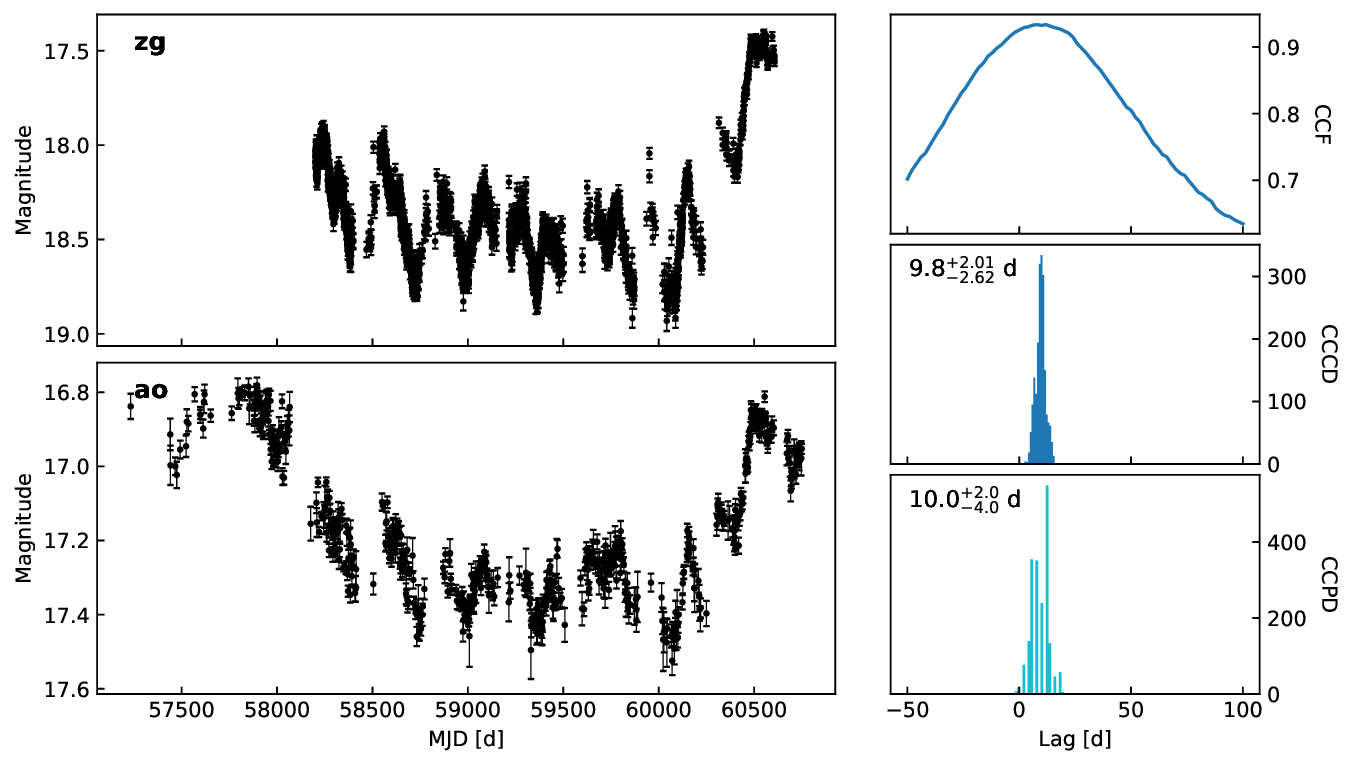}      \\[2pt]
		\includegraphics[width=0.45\textwidth]{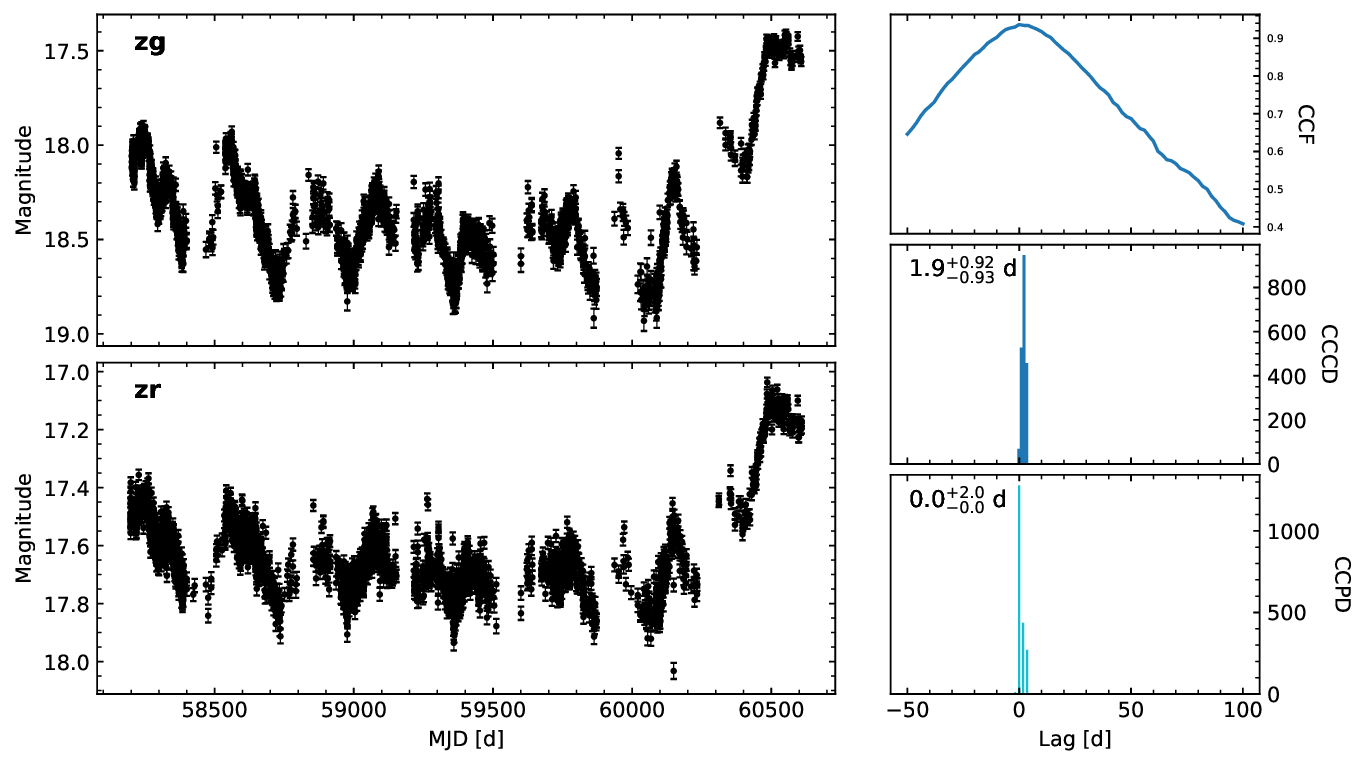}      &
		\includegraphics[width=0.45\textwidth]{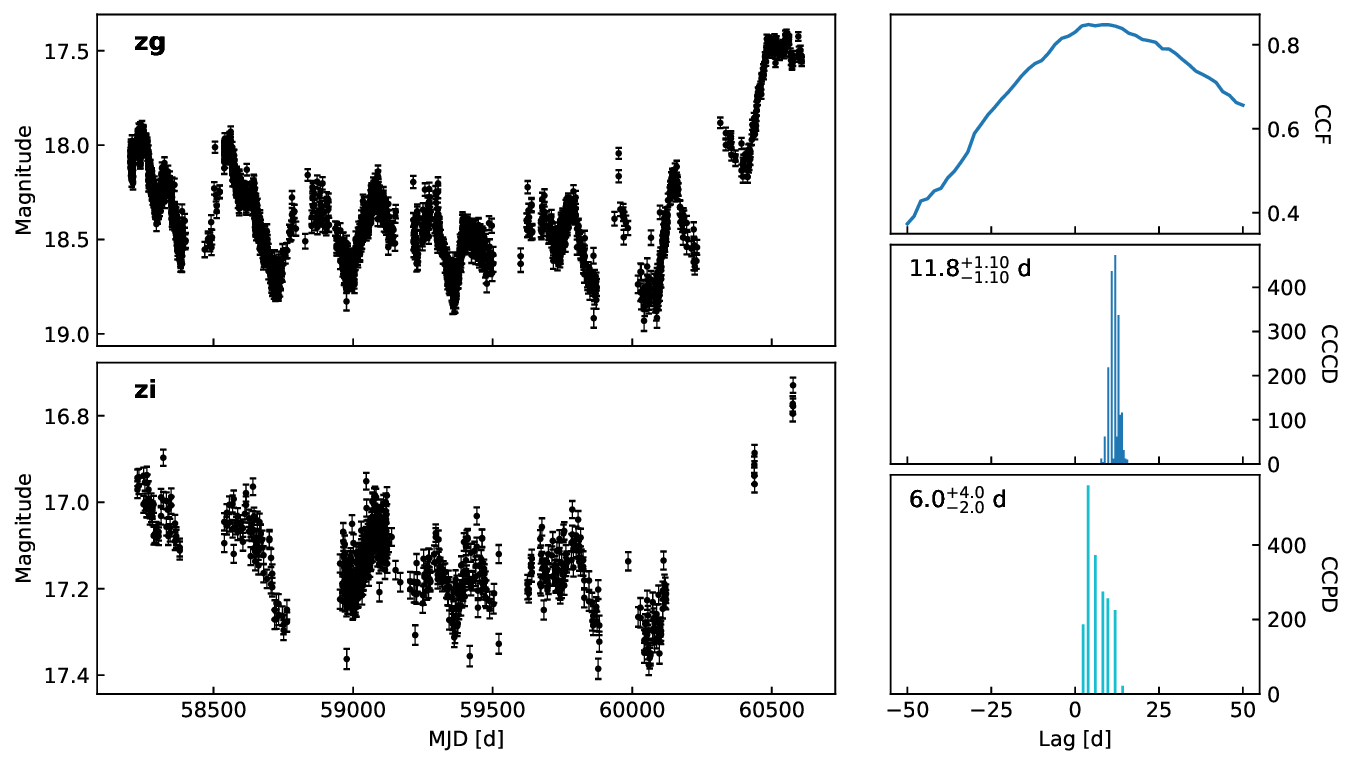}      \\
	\end{tabular}
	\caption{Cross-correlation analysis of different LC pairs, which
	were used to estimate inter-band time lags. Each set of the panels 
	displays the two LCs being cross-correlated, the resulting CCF, 
	and the CCCD/CCPD lag distributions from \texttt{PyCCF}.}
	\label{fig:multi-lag}
\end{figure}

In Figure~\ref{fig:multi-lag}, individual cross-correlation analysis 
plots are provided.

\bibliography{qpo}{}

\begin{thebibliography}{}
\expandafter\ifx\csname natexlab\endcsname\relax\def\natexlab#1{#1}\fi

\bibitem[{{Abazajian} {et~al.}(2009){Abazajian}, {Adelman-McCarthy},
  {Ag{\"u}eros}, {Allam}, {Allende Prieto}, {An}, {Anderson}, {Anderson},
  {Annis}, {Bahcall}, {Bailer-Jones}, {Barentine}, {Bassett}, {Becker},
  {Beers}, {Bell}, {Belokurov}, {Berlind}, {Berman}, {Bernardi}, {Bickerton},
  {Bizyaev}, {Blakeslee}, {Blanton}, {Bochanski}, {Boroski}, {Brewington},
  {Brinchmann}, {Brinkmann}, {Brunner}, {Budav{\'a}ri}, {Carey}, {Carliles},
  {Carr}, {Castander}, {Cinabro}, {Connolly}, {Csabai}, {Cunha}, {Czarapata},
  {Davenport}, {de Haas}, {Dilday}, {Doi}, {Eisenstein}, {Evans}, {Evans},
  {Fan}, {Friedman}, {Frieman}, {Fukugita}, {G{\"a}nsicke}, {Gates},
  {Gillespie}, {Gilmore}, {Gonzalez}, {Gonzalez}, {Grebel}, {Gunn},
  {Gy{\"o}ry}, {Hall}, {Harding}, {Harris}, {Harvanek}, {Hawley}, {Hayes},
  {Heckman}, {Hendry}, {Hennessy}, {Hindsley}, {Hoblitt}, {Hogan}, {Hogg},
  {Holtzman}, {Hyde}, {Ichikawa}, {Ichikawa}, {Im}, {Ivezi{\'c}}, {Jester},
  {Jiang}, {Johnson}, {Jorgensen}, {Juri{\'c}}, {Kent}, {Kessler}, {Kleinman},
  {Knapp}, {Konishi}, {Kron}, {Krzesinski}, {Kuropatkin}, {Lampeitl},
  {Lebedeva}, {Lee}, {Lee}, {French Leger}, {L{\'e}pine}, {Li}, {Lima}, {Lin},
  {Long}, {Loomis}, {Loveday}, {Lupton}, {Magnier}, {Malanushenko},
  {Malanushenko}, {Mandelbaum}, {Margon}, {Marriner}, {Mart{\'\i}nez-Delgado},
  {Matsubara}, {McGehee}, {McKay}, {Meiksin}, {Morrison}, {Mullally}, {Munn},
  {Murphy}, {Nash}, {Nebot}, {Neilsen}, {Newberg}, {Newman}, {Nichol},
  {Nicinski}, {Nieto-Santisteban}, {Nitta}, {Okamura}, {Oravetz}, {Ostriker},
  {Owen}, {Padmanabhan}, {Pan}, {Park}, {Pauls}, {Peoples}, {Percival}, {Pier},
  {Pope}, {Pourbaix}, {Price}, {Purger}, {Quinn}, {Raddick}, {Re Fiorentin},
  {Richards}, {Richmond}, {Riess}, {Rix}, {Rockosi}, {Sako}, {Schlegel},
  {Schneider}, {Scholz}, {Schreiber}, {Schwope}, {Seljak}, {Sesar}, {Sheldon},
  {Shimasaku}, {Sibley}, {Simmons}, {Sivarani}, {Allyn Smith}, {Smith},
  {Smol{\v{c}}i{\'c}}, {Snedden}, {Stebbins}, {Steinmetz}, {Stoughton},
  {Strauss}, {SubbaRao}, {Suto}, {Szalay}, {Szapudi}, {Szkody}, {Tanaka},
  {Tegmark}, {Teodoro}, {Thakar}, {Tremonti}, {Tucker}, {Uomoto}, {Vanden
  Berk}, {Vandenberg}, {Vidrih}, {Vogeley}, {Voges}, {Vogt}, {Wadadekar},
  {Watters}, {Weinberg}, {West}, {White}, {Wilhite}, {Wonders}, {Yanny},
  {Yocum}, {York}, {Zehavi}, {Zibetti}, \& {Zucker}}]{aaa+09}
{Abazajian}, K.~N., {Adelman-McCarthy}, J.~K., {Ag{\"u}eros}, M.~A., {et~al.}
  2009, \apjs, 182, 543

\bibitem[{{Abramowicz} {et~al.}(2004){Abramowicz}, {Klu{\'z}niak},
  {McClintock}, \& {Remillard}}]{abr+04}
{Abramowicz}, M.~A., {Klu{\'z}niak}, W., {McClintock}, J.~E., \& {Remillard},
  R.~A. 2004, \apjl, 609, L63

\bibitem[{{Alston} {et~al.}(2014){Alston}, {Markeviciute}, {Kara}, {Fabian}, \&
  {Middleton}}]{als+14}
{Alston}, W.~N., {Markeviciute}, J., {Kara}, E., {Fabian}, A.~C., \&
  {Middleton}, M. 2014, \mnras, 445, L16

\bibitem[{{Alston} {et~al.}(2015){Alston}, {Parker},
  {Markevi{\v{c}}i{\={u}}t{\.{e}}}, {Fabian}, {Middleton}, {Lohfink}, {Kara},
  \& {Pinto}}]{als+15}
{Alston}, W.~N., {Parker}, M.~L., {Markevi{\v{c}}i{\={u}}t{\.{e}}}, J.,
  {et~al.} 2015, \mnras, 449, 467

\bibitem[{Aydin(2017)}]{wwzcode}
Aydin, M.~E. 2017, {eaydin/WWZ: v1.0.0},
  \url{http://doi.org/10.5281/zenodo.375648},  Zenodo,
  doi:10.5281/zenodo.375648

\bibitem[{{Bardeen} {et~al.}(1972){Bardeen}, {Press}, \& {Teukolsky}}]{bpt72}
{Bardeen}, J.~M., {Press}, W.~H., \& {Teukolsky}, S.~A. 1972, \apj, 178, 347

\bibitem[{{Bellm} {et~al.}(2019){Bellm}, {Kulkarni}, {Graham}, {Dekany},
  {Smith}, {Riddle}, {Masci}, {Helou}, {Prince}, {Adams}, {Barbarino},
  {Barlow}, {Bauer}, {Beck}, {Belicki}, {Biswas}, {Blagorodnova}, {Bodewits},
  {Bolin}, {Brinnel}, {Brooke}, {Bue}, {Bulla}, {Burruss}, {Cenko}, {Chang},
  {Connolly}, {Coughlin}, {Cromer}, {Cunningham}, {De}, {Delacroix}, {Desai},
  {Duev}, {Eadie}, {Farnham}, {Feeney}, {Feindt}, {Flynn}, {Franckowiak},
  {Frederick}, {Fremling}, {Gal-Yam}, {Gezari}, {Giomi}, {Goldstein},
  {Golkhou}, {Goobar}, {Groom}, {Hacopians}, {Hale}, {Henning}, {Ho}, {Hover},
  {Howell}, {Hung}, {Huppenkothen}, {Imel}, {Ip}, {Ivezi{\'c}}, {Jackson},
  {Jones}, {Juric}, {Kasliwal}, {Kaspi}, {Kaye}, {Kelley}, {Kowalski},
  {Kramer}, {Kupfer}, {Landry}, {Laher}, {Lee}, {Lin}, {Lin}, {Lunnan},
  {Giomi}, {Mahabal}, {Mao}, {Miller}, {Monkewitz}, {Murphy}, {Ngeow},
  {Nordin}, {Nugent}, {Ofek}, {Patterson}, {Penprase}, {Porter}, {Rauch},
  {Rebbapragada}, {Reiley}, {Rigault}, {Rodriguez}, {van Roestel}, {Rusholme},
  {van Santen}, {Schulze}, {Shupe}, {Singer}, {Soumagnac}, {Stein}, {Surace},
  {Sollerman}, {Szkody}, {Taddia}, {Terek}, {Van Sistine}, {van Velzen},
  {Vestrand}, {Walters}, {Ward}, {Ye}, {Yu}, {Yan}, \& {Zolkower}}]{bkg+19}
{Bellm}, E.~C., {Kulkarni}, S.~R., {Graham}, M.~J., {et~al.} 2019, \pasp, 131,
  018002

\bibitem[{{Burrows} {et~al.}(2005){Burrows}, {Hill}, {Nousek}, {Kennea},
  {Wells}, {Osborne}, {Abbey}, {Beardmore}, {Mukerjee}, {Short}, {Chincarini},
  {Campana}, {Citterio}, {Moretti}, {Pagani}, {Tagliaferri}, {Giommi},
  {Capalbi}, {Tamburelli}, {Angelini}, {Cusumano}, {Br{\"a}uninger}, {Burkert},
  \& {Hartner}}]{xrt05}
{Burrows}, D.~N., {Hill}, J.~E., {Nousek}, J.~A., {et~al.} 2005, \ssr, 120, 165

\bibitem[{{Colpi}(2014)}]{col14}
{Colpi}, M. 2014, \ssr, 183, 189

\bibitem[{{Cowperthwaite} \& {Reynolds}(2012)}]{cr12}
{Cowperthwaite}, P.~S., \& {Reynolds}, C.~S. 2012, \apjl, 752, L21

\bibitem[{{Cowsik} {et~al.}(2002){Cowsik}, {Srinivasan}, \& {Prabhu}}]{csp02}
{Cowsik}, R., {Srinivasan}, R., \& {Prabhu}, T.~P. 2002, Bulletin of the
  Astronomical Society of India, 30, 105

\bibitem[{{Cui} {et~al.}(2012){Cui}, {Zhao}, {Chu}, {Li}, {Li}, {Zhang}, {Su},
  {Yao}, {Wang}, {Xing}, {Li}, {Zhu}, {Wang}, {Gu}, {Luo}, {Xu}, {Zhang},
  {Liu}, {Zhang}, {Yang}, {Cao}, {Chen}, {Chen}, {Chen}, {Chen}, {Chu}, {Feng},
  {Gong}, {Hou}, {Hu}, {Hu}, {Hu}, {Jia}, {Jiang}, {Jiang}, {Jiang}, {Jin},
  {Li}, {Li}, {Li}, {Liu}, {Liu}, {Lu}, {Mao}, {Men}, {Qi}, {Qi}, {Shi},
  {Tang}, {Tao}, {Wang}, {Wang}, {Wang}, {Wang}, {Wang}, {Wang}, {Wang},
  {Wang}, {Wang}, {Wang}, {Wang}, {Wang}, {Xu}, {Xu}, {Yang}, {Yu}, {Yuan},
  {Yuan}, {Zhai}, {Zhang}, {Zhang}, {Zhang}, {Zhao}, {Zhou}, {Zhou}, {Zhu}, \&
  {Zou}}]{lamost12}
{Cui}, X.-Q., {Zhao}, Y.-H., {Chu}, Y.-Q., {et~al.} 2012, Research in Astronomy
  and Astrophysics, 12, 1197

\bibitem[{{Davis} \& {Laor}(2011)}]{dl+11}
{Davis}, S.~W., \& {Laor}, A. 2011, \apj, 728, 98

\bibitem[{{DESI Collaboration} {et~al.}(2016){DESI Collaboration}, {Aghamousa},
  {Aguilar}, {Ahlen}, {Alam}, {Allen}, {Allende Prieto}, {Annis}, {Bailey},
  {Balland}, {Ballester}, {Baltay}, {Beaufore}, {Bebek}, {Beers}, {Bell},
  {Bernal}, {Besuner}, {Beutler}, {Blake}, {Bleuler}, {Blomqvist}, {Blum},
  {Bolton}, {Briceno}, {Brooks}, {Brownstein}, {Buckley-Geer}, {Burden},
  {Burtin}, {Busca}, {Cahn}, {Cai}, {Cardiel-Sas}, {Carlberg}, {Carton},
  {Casas}, {Castander}, {Cervantes-Cota}, {Claybaugh}, {Close}, {Coker},
  {Cole}, {Comparat}, {Cooper}, {Cousinou}, {Crocce}, {Cuby}, {Cunningham},
  {Davis}, {Dawson}, {de la Macorra}, {De Vicente}, {Delubac}, {Derwent},
  {Dey}, {Dhungana}, {Ding}, {Doel}, {Duan}, {Ealet}, {Edelstein},
  {Eftekharzadeh}, {Eisenstein}, {Elliott}, {Escoffier}, {Evatt}, {Fagrelius},
  {Fan}, {Fanning}, {Farahi}, {Farihi}, {Favole}, {Feng}, {Fernandez},
  {Findlay}, {Finkbeiner}, {Fitzpatrick}, {Flaugher}, {Flender}, {Font-Ribera},
  {Forero-Romero}, {Fosalba}, {Frenk}, {Fumagalli}, {Gaensicke}, {Gallo},
  {Garcia-Bellido}, {Gaztanaga}, {Pietro Gentile Fusillo}, {Gerard},
  {Gershkovich}, {Giannantonio}, {Gillet}, {Gonzalez-de-Rivera},
  {Gonzalez-Perez}, {Gott}, {Graur}, {Gutierrez}, {Guy}, {Habib}, {Heetderks},
  {Heetderks}, {Heitmann}, {Hellwing}, {Herrera}, {Ho}, {Holland}, {Honscheid},
  {Huff}, {Hutchinson}, {Huterer}, {Hwang}, {Illa Laguna}, {Ishikawa},
  {Jacobs}, {Jeffrey}, {Jelinsky}, {Jennings}, {Jiang}, {Jimenez}, {Johnson},
  {Joyce}, {Jullo}, {Juneau}, {Kama}, {Karcher}, {Karkar}, {Kehoe}, {Kennamer},
  {Kent}, {Kilbinger}, {Kim}, {Kirkby}, {Kisner}, {Kitanidis}, {Kneib},
  {Koposov}, {Kovacs}, {Koyama}, {Kremin}, {Kron}, {Kronig}, {Kueter-Young},
  {Lacey}, {Lafever}, {Lahav}, {Lambert}, {Lampton}, {Landriau}, {Lang},
  {Lauer}, {Le Goff}, {Le Guillou}, {Le Van Suu}, {Lee}, {Lee}, {Leitner},
  {Lesser}, {Levi}, {L'Huillier}, {Li}, {Liang}, {Lin}, {Linder}, {Loebman},
  {Luki{\'c}}, {Ma}, {MacCrann}, {Magneville}, {Makarem}, {Manera}, {Manser},
  {Marshall}, {Martini}, {Massey}, {Matheson}, {McCauley}, {McDonald},
  {McGreer}, {Meisner}, {Metcalfe}, {Miller}, {Miquel}, {Moustakas}, {Myers},
  {Naik}, {Newman}, {Nichol}, {Nicola}, {Nicolati da Costa}, {Nie}, {Niz},
  {Norberg}, {Nord}, {Norman}, {Nugent}, {O'Brien}, {Oh}, \& {Olsen}}]{desi16}
{DESI Collaboration}, {Aghamousa}, A., {Aguilar}, J., {et~al.} 2016, arXiv
  e-prints, arXiv:1611.00036

\bibitem[{{Drake} {et~al.}(2009){Drake}, {Djorgovski}, {Mahabal}, {Beshore},
  {Larson}, {Graham}, {Williams}, {Christensen}, {Catelan}, {Boattini},
  {Gibbs}, {Hill}, \& {Kowalski}}]{ddm+09}
{Drake}, A.~J., {Djorgovski}, S.~G., {Mahabal}, A., {et~al.} 2009, \apj, 696,
  870

\bibitem[{{Emmanoulopoulos} {et~al.}(2013){Emmanoulopoulos}, {McHardy}, \&
  {Papadakis}}]{emp13}
{Emmanoulopoulos}, D., {McHardy}, I.~M., \& {Papadakis}, I.~E. 2013, \mnras,
  433, 907

\bibitem[{{Evans} {et~al.}(2009){Evans}, {Beardmore}, {Page}, {Osborne},
  {O'Brien}, {Willingale}, {Starling}, {Burrows}, {Godet}, {Vetere}, {Racusin},
  {Goad}, {Wiersema}, {Angelini}, {Capalbi}, {Chincarini}, {Gehrels}, {Kennea},
  {Margutti}, {Morris}, {Mountford}, {Pagani}, {Perri}, {Romano}, \&
  {Tanvir}}]{ebp+09}
{Evans}, P.~A., {Beardmore}, A.~P., {Page}, K.~L., {et~al.} 2009, \mnras, 397,
  1177

\bibitem[{{Fausnaugh} {et~al.}(2016){Fausnaugh}, {Denney}, {Barth}, {Bentz},
  {Bottorff}, {Carini}, {Croxall}, {De Rosa}, {Goad}, {Horne}, {Joner},
  {Kaspi}, {Kim}, {Klimanov}, {Kochanek}, {Leonard}, {Netzer}, {Peterson},
  {Schn{\"u}lle}, {Sergeev}, {Vestergaard}, {Zheng}, {Zu}, {Anderson},
  {Ar{\'e}valo}, {Bazhaw}, {Borman}, {Boroson}, {Brandt}, {Breeveld}, {Brewer},
  {Cackett}, {Crenshaw}, {Dalla Bont{\`a}}, {De Lorenzo-C{\'a}ceres},
  {Dietrich}, {Edelson}, {Efimova}, {Ely}, {Evans}, {Filippenko}, {Flatland},
  {Gehrels}, {Geier}, {Gelbord}, {Gonzalez}, {Gorjian}, {Grier}, {Grupe},
  {Hall}, {Hicks}, {Horenstein}, {Hutchison}, {Im}, {Jensen}, {Jones},
  {Kaastra}, {Kelly}, {Kennea}, {Kim}, {Korista}, {Kriss}, {Lee}, {Lira},
  {MacInnis}, {Manne-Nicholas}, {Mathur}, {McHardy}, {Montouri}, {Musso},
  {Nazarov}, {Norris}, {Nousek}, {Okhmat}, {Pancoast}, {Papadakis}, {Parks},
  {Pei}, {Pogge}, {Pott}, {Rafter}, {Rix}, {Saylor}, {Schimoia}, {Siegel},
  {Spencer}, {Starkey}, {Sung}, {Teems}, {Treu}, {Turner}, {Uttley},
  {Villforth}, {Weiss}, {Woo}, {Yan}, \& {Young}}]{fdb+16}
{Fausnaugh}, M.~M., {Denney}, K.~D., {Barth}, A.~J., {et~al.} 2016, \apj, 821,
  56

\bibitem[{{Foster}(1996)}]{wwz96}
{Foster}, G. 1996, \aj, 112, 1709

\bibitem[{{Gaia Collaboration} {et~al.}(2016){Gaia Collaboration}, {Prusti},
  {de Bruijne}, {Brown}, {Vallenari}, {Babusiaux}, {Bailer-Jones}, {Bastian},
  {Biermann}, {Evans}, {Eyer}, {Jansen}, {Jordi}, {Klioner}, {Lammers},
  {Lindegren}, {Luri}, {Mignard}, {Milligan}, {Panem}, {Poinsignon},
  {Pourbaix}, {Randich}, {Sarri}, {Sartoretti}, {Siddiqui}, {Soubiran},
  {Valette}, {van Leeuwen}, {Walton}, {Aerts}, {Arenou}, {Cropper}, {Drimmel},
  {H{\o}g}, {Katz}, {Lattanzi}, {O'Mullane}, {Grebel}, {Holland}, {Huc},
  {Passot}, {Bramante}, {Cacciari}, {Casta{\~n}eda}, {Chaoul}, {Cheek}, {De
  Angeli}, {Fabricius}, {Guerra}, {Hern{\'a}ndez}, {Jean-Antoine-Piccolo},
  {Masana}, {Messineo}, {Mowlavi}, {Nienartowicz}, {Ord{\'o}{\~n}ez-Blanco},
  {Panuzzo}, {Portell}, {Richards}, {Riello}, {Seabroke}, {Tanga},
  {Th{\'e}venin}, {Torra}, {Els}, {Gracia-Abril}, {Comoretto},
  {Garcia-Reinaldos}, {Lock}, {Mercier}, {Altmann}, {Andrae}, {Astraatmadja},
  {Bellas-Velidis}, {Benson}, {Berthier}, {Blomme}, {Busso}, {Carry},
  {Cellino}, {Clementini}, {Cowell}, {Creevey}, {Cuypers}, {Davidson}, {De
  Ridder}, {de Torres}, {Delchambre}, {Dell'Oro}, {Ducourant}, {Fr{\'e}mat},
  {Garc{\'\i}a-Torres}, {Gosset}, {Halbwachs}, {Hambly}, {Harrison}, {Hauser},
  {Hestroffer}, {Hodgkin}, {Huckle}, {Hutton}, {Jasniewicz}, {Jordan},
  {Kontizas}, {Korn}, {Lanzafame}, {Manteiga}, {Moitinho}, {Muinonen},
  {Osinde}, {Pancino}, {Pauwels}, {Petit}, {Recio-Blanco}, {Robin}, {Sarro},
  {Siopis}, {Smith}, {Smith}, {Sozzetti}, {Thuillot}, {van Reeven}, {Viala},
  {Abbas}, {Abreu Aramburu}, {Accart}, {Aguado}, {Allan}, {Allasia},
  {Altavilla}, {{\'A}lvarez}, {Alves}, {Anderson}, {Andrei}, {Anglada Varela},
  {Antiche}, {Antoja}, {Ant{\'o}n}, {Arcay}, {Atzei}, {Ayache}, {Bach},
  {Baker}, {Balaguer-N{\'u}{\~n}ez}, {Barache}, {Barata}, {Barbier}, {Barblan},
  {Baroni}, {Barrado y Navascu{\'e}s}, {Barros}, {Barstow}, {Becciani},
  {Bellazzini}, {Bellei}, {Bello Garc{\'\i}a}, {Belokurov}, {Bendjoya},
  {Berihuete}, {Bianchi}, {Bienaym{\'e}}, {Billebaud}, {Blagorodnova},
  {Blanco-Cuaresma}, {Boch}, {Bombrun}, {Borrachero}, {Bouquillon}, {Bourda},
  {Bouy}, {Bragaglia}, {Breddels}, {Brouillet}, {Br{\"u}semeister},
  {Bucciarelli}, {Budnik}, {Burgess}, {Burgon}, {Burlacu}, {Busonero}, {Buzzi},
  {Caffau}, {Cambras}, {Campbell}, {Cancelliere}, {Cantat-Gaudin}, {Carlucci},
  {Carrasco}, {Castellani}, {Charlot}, {Charnas}, {Charvet}, {Chassat},
  {Chiavassa}, {Clotet}, {Cocozza}, {Collins}, {Collins}, {Costigan}, {Crifo},
  {Cross}, {Crosta}, {Crowley}, {Dafonte}, {Damerdji}, {Dapergolas}, {David},
  {David}, {De Cat}, {de Felice}, {de Laverny}, {De Luise}, {De March}, {de
  Martino}, {de Souza}, {Debosscher}, {del Pozo}, {Delbo}, {Delgado},
  {Delgado}, {di Marco}, {Di Matteo}, {Diakite}, {Distefano}, {Dolding}, {Dos
  Anjos}, {Drazinos}, {Dur{\'a}n}, {Dzigan}, {Ecale}, {Edvardsson}, {Enke},
  {Erdmann}, {Escolar}, {Espina}, {Evans}, {Eynard Bontemps}, {Fabre},
  {Fabrizio}, {Faigler}, {Falc{\~a}o}, {Farr{\`a}s Casas}, {Faye}, {Federici},
  {Fedorets}, {Fern{\'a}ndez-Hern{\'a}ndez}, {Fernique}, {Fienga}, {Figueras},
  {Filippi}, {Findeisen}, {Fonti}, {Fouesneau}, {Fraile}, {Fraser}, {Fuchs},
  {Furnell}, {Gai}, {Galleti}, {Galluccio}, {Garabato}, {Garc{\'\i}a-Sedano},
  {Gar{\'e}}, {Garofalo}, {Garralda}, {Gavras}, {Gerssen}, {Geyer}, {Gilmore},
  {Girona}, {Giuffrida}, {Gomes}, {Gonz{\'a}lez-Marcos},
  {Gonz{\'a}lez-N{\'u}{\~n}ez}, {Gonz{\'a}lez-Vidal}, {Granvik}, {Guerrier},
  {Guillout}, {Guiraud}, {G{\'u}rpide}, {Guti{\'e}rrez-S{\'a}nchez}, {Guy},
  {Haigron}, {Hatzidimitriou}, {Haywood}, {Heiter}, {Helmi}, {Hobbs},
  {Hofmann}, {Holl}, {Holland}, {Hunt}, {Hypki}, {Icardi}, {Irwin}, {Jevardat
  de Fombelle}, {Jofr{\'e}}, {Jonker}, {Jorissen}, {Julbe}, {Karampelas},
  {Kochoska}, {Kohley}, {Kolenberg}, {Kontizas}, {Koposov}, {Kordopatis},
  {Koubsky}, {Kowalczyk}, {Krone-Martins}, {Kudryashova}, {Kull}, {Bachchan},
  {Lacoste-Seris}, {Lanza}, {Lavigne}, {Le Poncin-Lafitte}, {Lebreton},
  {Lebzelter}, {Leccia}, {Leclerc}, {Lecoeur-Taibi}, {Lemaitre}, {Lenhardt},
  {Leroux}, {Liao}, {Licata}, {Lindstr{\o}m}, {Lister}, {Livanou}, {Lobel},
  {L{\"o}ffler}, {L{\'o}pez}, {Lopez-Lozano}, {Lorenz}, {Loureiro},
  {MacDonald}, {Magalh{\~a}es Fernandes}, {Managau}, {Mann}, {Mantelet},
  {Marchal}, {Marchant}, {Marconi}, {Marie}, {Marinoni}, {Marrese},
  {Marschalk{\'o}}, {Marshall}, {Mart{\'\i}n-Fleitas}, {Martino}, {Mary},
  {Matijevi{\v{c}}}, {Mazeh}, {McMillan}, {Messina}, {Mestre}, {Michalik},
  {Millar}, {Miranda}, {Molina}, {Molinaro}, {Molinaro}, {Moln{\'a}r},
  {Moniez}, {Montegriffo}, {Monteiro}, {Mor}, {Mora}, {Morbidelli}, {Morel},
  {Morgenthaler}, {Morley}, {Morris}, {Mulone}, {Muraveva}, {Musella},
  {Narbonne}, {Nelemans}, {Nicastro}, {Noval}, {Ord{\'e}novic},
  {Ordieres-Mer{\'e}}, {Osborne}, {Pagani}, {Pagano}, {Pailler}, {Palacin},
  {Palaversa}, {Parsons}, {Paulsen}, {Pecoraro}, {Pedrosa}, {Pentik{\"a}inen},
  {Pereira}, {Pichon}, {Piersimoni}, {Pineau}, {Plachy}, {Plum}, {Poujoulet},
  {Pr{\v{s}}a}, {Pulone}, {Ragaini}, {Rago}, {Rambaux}, {Ramos-Lerate},
  {Ranalli}, {Rauw}, {Read}, {Regibo}, {Renk}, {Reyl{\'e}}, {Ribeiro},
  {Rimoldini}, {Ripepi}, {Riva}, {Rixon}, {Roelens}, {Romero-G{\'o}mez},
  {Rowell}, {Royer}, {Rudolph}, {Ruiz-Dern}, {Sadowski}, {Sagrist{\`a}
  Sell{\'e}s}, {Sahlmann}, {Salgado}, {Salguero}, {Sarasso}, {Savietto},
  {Schnorhk}, {Schultheis}, {Sciacca}, {Segol}, {Segovia}, {Segransan},
  {Serpell}, {Shih}, {Smareglia}, {Smart}, {Smith}, {Solano}, {Solitro},
  {Sordo}, {Soria Nieto}, {Souchay}, {Spagna}, {Spoto}, {Stampa}, {Steele},
  {Steidelm{\"u}ller}, {Stephenson}, {Stoev}, {Suess}, {S{\"u}veges}, {Surdej},
  {Szabados}, {Szegedi-Elek}, {Tapiador}, {Taris}, {Tauran}, {Taylor},
  {Teixeira}, {Terrett}, {Tingley}, {Trager}, {Turon}, {Ulla}, {Utrilla},
  {Valentini}, {van Elteren}, {Van Hemelryck}, {van Leeuwen}, {Varadi},
  {Vecchiato}, {Veljanoski}, {Via}, {Vicente}, {Vogt}, {Voss}, {Votruba},
  {Voutsinas}, {Walmsley}, {Weiler}, {Weingrill}, {Werner}, {Wevers},
  {Whitehead}, {Wyrzykowski}, {Yoldas}, {{\v{Z}}erjal}, {Zucker}, {Zurbach},
  {Zwitter}, {Alecu}, {Allen}, {Allende Prieto}, {Amorim},
  {Anglada-Escud{\'e}}, {Arsenijevic}, {Azaz}, {Balm}, {Beck}, {Bernstein},
  {Bigot}, {Bijaoui}, {Blasco}, {Bonfigli}, {Bono}, {Boudreault}, {Bressan},
  {Brown}, {Brunet}, {Bunclark}, {Buonanno}, {Butkevich}, {Carret}, {Carrion},
  {Chemin}, {Ch{\'e}reau}, {Corcione}, {Darmigny}, {de Boer}, {de Teodoro}, {de
  Zeeuw}, {Delle Luche}, {Domingues}, {Dubath}, {Fodor}, {Fr{\'e}zouls},
  {Fries}, {Fustes}, {Fyfe}, {Gallardo}, {Gallegos}, {Gardiol}, {Gebran},
  {Gomboc}, {G{\'o}mez}, {Grux}, {Gueguen}, {Heyrovsky}, {Hoar}, {Iannicola},
  {Isasi Parache}, {Janotto}, {Joliet}, {Jonckheere}, {Keil}, {Kim},
  {Klagyivik}, {Klar}, {Knude}, {Kochukhov}, {Kolka}, {Kos}, {Kutka}, {Lainey},
  {LeBouquin}, {Liu}, {Loreggia}, {Makarov}, {Marseille}, {Martayan},
  {Martinez-Rubi}, {Massart}, {Meynadier}, {Mignot}, {Munari}, {Nguyen},
  {Nordlander}, {Ocvirk}, {O'Flaherty}, {Olias Sanz}, {Ortiz}, {Osorio},
  {Oszkiewicz}, {Ouzounis}, {Palmer}, {Park}, {Pasquato}, {Peltzer}, {Peralta},
  {P{\'e}turaud}, {Pieniluoma}, {Pigozzi}, {Poels}, {Prat}, {Prod'homme},
  {Raison}, {Rebordao}, {Risquez}, {Rocca-Volmerange}, {Rosen}, {Ruiz-Fuertes},
  {Russo}, {Sembay}, {Serraller Vizcaino}, {Short}, {Siebert}, {Silva},
  {Sinachopoulos}, {Slezak}, {Soffel}, {Sosnowska}, {Strai{\v{z}}ys}, {ter
  Linden}, {Terrell}, {Theil}, {Tiede}, {Troisi}, {Tsalmantza}, {Tur},
  {Vaccari}, {Vachier}, {Valles}, {Van Hamme}, {Veltz}, {Virtanen}, {Wallut},
  {Wichmann}, {Wilkinson}, {Ziaeepour}, \& {Zschocke}}]{gaia16}
{Gaia Collaboration}, {Prusti}, T., {de Bruijne}, J.~H.~J., {et~al.} 2016,
  \aap, 595, A1

\bibitem[{{Gaia Collaboration} {et~al.}(2023){Gaia Collaboration}, {Vallenari},
  {Brown}, {Prusti}, {de Bruijne}, {Arenou}, {Babusiaux}, {Biermann},
  {Creevey}, {Ducourant}, {Evans}, {Eyer}, {Guerra}, {Hutton}, {Jordi},
  {Klioner}, {Lammers}, {Lindegren}, {Luri}, {Mignard}, {Panem}, {Pourbaix},
  {Randich}, {Sartoretti}, {Soubiran}, {Tanga}, {Walton}, {Bailer-Jones},
  {Bastian}, {Drimmel}, {Jansen}, {Katz}, {Lattanzi}, {van Leeuwen}, {Bakker},
  {Cacciari}, {Casta{\~n}eda}, {De Angeli}, {Fabricius}, {Fouesneau},
  {Fr{\'e}mat}, {Galluccio}, {Guerrier}, {Heiter}, {Masana}, {Messineo},
  {Mowlavi}, {Nicolas}, {Nienartowicz}, {Pailler}, {Panuzzo}, {Riclet}, {Roux},
  {Seabroke}, {Sordo}, {Th{\'e}venin}, {Gracia-Abril}, {Portell}, {Teyssier},
  {Altmann}, {Andrae}, {Audard}, {Bellas-Velidis}, {Benson}, {Berthier},
  {Blomme}, {Burgess}, {Busonero}, {Busso}, {C{\'a}novas}, {Carry}, {Cellino},
  {Cheek}, {Clementini}, {Damerdji}, {Davidson}, {de Teodoro}, {Nu{\~n}ez
  Campos}, {Delchambre}, {Dell'Oro}, {Esquej}, {Fern{\'a}ndez-Hern{\'a}ndez},
  {Fraile}, {Garabato}, {Garc{\'\i}a-Lario}, {Gosset}, {Haigron}, {Halbwachs},
  {Hambly}, {Harrison}, {Hern{\'a}ndez}, {Hestroffer}, {Hodgkin}, {Holl},
  {Jan{\ss}en}, {Jevardat de Fombelle}, {Jordan}, {Krone-Martins}, {Lanzafame},
  {L{\"o}ffler}, {Marchal}, {Marrese}, {Moitinho}, {Muinonen}, {Osborne},
  {Pancino}, {Pauwels}, {Recio-Blanco}, {Reyl{\'e}}, {Riello}, {Rimoldini},
  {Roegiers}, {Rybizki}, {Sarro}, {Siopis}, {Smith}, {Sozzetti}, {Utrilla},
  {van Leeuwen}, {Abbas}, {{\'A}brah{\'a}m}, {Abreu Aramburu}, {Aerts},
  {Aguado}, {Ajaj}, {Aldea-Montero}, {Altavilla}, {{\'A}lvarez}, {Alves},
  {Anders}, {Anderson}, {Anglada Varela}, {Antoja}, {Baines}, {Baker},
  {Balaguer-N{\'u}{\~n}ez}, {Balbinot}, {Balog}, {Barache}, {Barbato},
  {Barros}, {Barstow}, {Bartolom{\'e}}, {Bassilana}, {Bauchet}, {Becciani},
  {Bellazzini}, {Berihuete}, {Bernet}, {Bertone}, {Bianchi}, {Binnenfeld},
  {Blanco-Cuaresma}, {Blazere}, {Boch}, {Bombrun}, {Bossini}, {Bouquillon},
  {Bragaglia}, {Bramante}, {Breedt}, {Bressan}, {Brouillet}, {Brugaletta},
  {Bucciarelli}, {Burlacu}, {Butkevich}, {Buzzi}, {Caffau}, {Cancelliere},
  {Cantat-Gaudin}, {Carballo}, {Carlucci}, {Carnerero}, {Carrasco},
  {Casamiquela}, {Castellani}, {Castro-Ginard}, {Chaoul}, {Charlot}, {Chemin},
  {Chiaramida}, {Chiavassa}, {Chornay}, {Comoretto}, {Contursi}, {Cooper},
  {Cornez}, {Cowell}, {Crifo}, {Cropper}, {Crosta}, {Crowley}, {Dafonte},
  {Dapergolas}, {David}, {David}, {de Laverny}, {De Luise}, {De March}, {De
  Ridder}, {de Souza}, {de Torres}, {del Peloso}, {del Pozo}, {Delbo},
  {Delgado}, {Delisle}, {Demouchy}, {Dharmawardena}, {Di Matteo}, {Diakite},
  {Diener}, {Distefano}, {Dolding}, {Edvardsson}, {Enke}, {Fabre}, {Fabrizio},
  {Faigler}, {Fedorets}, {Fernique}, {Fienga}, {Figueras}, {Fournier},
  {Fouron}, {Fragkoudi}, {Gai}, {Garcia-Gutierrez}, {Garcia-Reinaldos},
  {Garc{\'\i}a-Torres}, {Garofalo}, {Gavel}, {Gavras}, {Gerlach}, {Geyer},
  {Giacobbe}, {Gilmore}, {Girona}, {Giuffrida}, {Gomel}, {Gomez},
  {Gonz{\'a}lez-N{\'u}{\~n}ez}, {Gonz{\'a}lez-Santamar{\'\i}a},
  {Gonz{\'a}lez-Vidal}, {Granvik}, {Guillout}, {Guiraud},
  {Guti{\'e}rrez-S{\'a}nchez}, {Guy}, {Hatzidimitriou}, {Hauser}, {Haywood},
  {Helmer}, {Helmi}, {Sarmiento}, {Hidalgo}, {Hilger}, {H{\l}adczuk}, {Hobbs},
  {Holland}, {Huckle}, {Jardine}, {Jasniewicz}, {Jean-Antoine Piccolo},
  {Jim{\'e}nez-Arranz}, {Jorissen}, {Juaristi Campillo}, {Julbe}, {Karbevska},
  {Kervella}, {Khanna}, {Kontizas}, {Kordopatis}, {Korn}, {K{\'o}sp{\'a}l},
  {Kostrzewa-Rutkowska}, {Kruszy{\'n}ska}, {Kun}, {Laizeau}, {Lambert},
  {Lanza}, {Lasne}, {Le Campion}, {Lebreton}, {Lebzelter}, {Leccia}, {Leclerc},
  {Lecoeur-Taibi}, {Liao}, {Licata}, {Lindstr{\o}m}, {Lister}, {Livanou},
  {Lobel}, {Lorca}, {Loup}, {Madrero Pardo}, {Magdaleno Romeo}, {Managau},
  {Mann}, {Manteiga}, {Marchant}, {Marconi}, {Marcos}, {Marcos Santos},
  {Mar{\'\i}n Pina}, {Marinoni}, {Marocco}, {Marshall}, {Martin Polo},
  {Mart{\'\i}n-Fleitas}, {Marton}, {Mary}, {Masip}, {Massari},
  {Mastrobuono-Battisti}, {Mazeh}, {McMillan}, {Messina}, {Michalik}, {Millar},
  {Mints}, {Molina}, {Molinaro}, {Moln{\'a}r}, {Monari}, {Mongui{\'o}},
  {Montegriffo}, {Montero}, {Mor}, {Mora}, {Morbidelli}, {Morel}, {Morris},
  {Muraveva}, {Murphy}, {Musella}, {Nagy}, {Noval}, {Oca{\~n}a}, {Ogden},
  {Ordenovic}, {Osinde}, {Pagani}, {Pagano}, {Palaversa}, {Palicio},
  {Pallas-Quintela}, {Panahi}, {Payne-Wardenaar}, {Pe{\~n}alosa Esteller},
  {Penttil{\"a}}, {Pichon}, {Piersimoni}, {Pineau}, {Plachy}, {Plum}, {Poggio},
  {Pr{\v{s}}a}, {Pulone}, {Racero}, {Ragaini}, {Rainer}, {Raiteri}, {Rambaux},
  {Ramos}, {Ramos-Lerate}, {Re Fiorentin}, {Regibo}, {Richards}, {Rios Diaz},
  {Ripepi}, {Riva}, {Rix}, {Rixon}, {Robichon}, {Robin}, {Robin}, {Roelens},
  {Rogues}, {Rohrbasser}, {Romero-G{\'o}mez}, {Rowell}, {Royer}, {Ruz Mieres},
  {Rybicki}, {Sadowski}, {S{\'a}ez N{\'u}{\~n}ez}, {Sagrist{\`a} Sell{\'e}s},
  {Sahlmann}, {Salguero}, {Samaras}, {Sanchez Gimenez}, {Sanna},
  {Santove{\~n}a}, {Sarasso}, {Schultheis}, {Sciacca}, {Segol}, {Segovia},
  {S{\'e}gransan}, {Semeux}, {Shahaf}, {Siddiqui}, {Siebert}, {Siltala},
  {Silvelo}, {Slezak}, {Slezak}, {Smart}, {Snaith}, {Solano}, {Solitro},
  {Souami}, {Souchay}, {Spagna}, {Spina}, {Spoto}, {Steele},
  {Steidelm{\"u}ller}, {Stephenson}, {S{\"u}veges}, {Surdej}, {Szabados},
  {Szegedi-Elek}, {Taris}, {Taylor}, {Teixeira}, {Tolomei}, {Tonello}, {Torra},
  {Torra}, {Torralba Elipe}, {Trabucchi}, {Tsounis}, {Turon}, {Ulla}, {Unger},
  {Vaillant}, {van Dillen}, {van Reeven}, {Vanel}, {Vecchiato}, {Viala},
  {Vicente}, {Voutsinas}, {Weiler}, {Wevers}, {Wyrzykowski}, {Yoldas}, {Yvard},
  {Zhao}, {Zorec}, {Zucker}, \& {Zwitter}}]{gaiaDR3}
{Gaia Collaboration}, {Vallenari}, A., {Brown}, A.~G.~A., {et~al.} 2023, \aap,
  674, A1

\bibitem[{{Gehrels} {et~al.}(2004){Gehrels}, {Chincarini}, {Giommi}, {Mason},
  {Nousek}, {Wells}, {White}, {Barthelmy}, {Burrows}, {Cominsky}, {Hurley},
  {Marshall}, {M{\'e}sz{\'a}ros}, {Roming}, {Angelini}, {Barbier}, {Belloni},
  {Campana}, {Caraveo}, {Chester}, {Citterio}, {Cline}, {Cropper}, {Cummings},
  {Dean}, {Feigelson}, {Fenimore}, {Frail}, {Fruchter}, {Garmire}, {Gendreau},
  {Ghisellini}, {Greiner}, {Hill}, {Hunsberger}, {Krimm}, {Kulkarni}, {Kumar},
  {Lebrun}, {Lloyd-Ronning}, {Markwardt}, {Mattson}, {Mushotzky}, {Norris},
  {Osborne}, {Paczynski}, {Palmer}, {Park}, {Parsons}, {Paul}, {Rees},
  {Reynolds}, {Rhoads}, {Sasseen}, {Schaefer}, {Short}, {Smale}, {Smith},
  {Stella}, {Tagliaferri}, {Takahashi}, {Tashiro}, {Townsley}, {Tueller},
  {Turner}, {Vietri}, {Voges}, {Ward}, {Willingale}, {Zerbi}, \&
  {Zhang}}]{swift04}
{Gehrels}, N., {Chincarini}, G., {Giommi}, P., {et~al.} 2004, \apj, 611, 1005

\bibitem[{{Gierli{\'n}ski} {et~al.}(2008){Gierli{\'n}ski}, {Middleton}, {Ward},
  \& {Done}}]{gmw+08}
{Gierli{\'n}ski}, M., {Middleton}, M., {Ward}, M., \& {Done}, C. 2008, \nat,
  455, 369

\bibitem[{{Graham} {et~al.}(2015){Graham}, {Djorgovski}, {Stern}, {Glikman},
  {Drake}, {Mahabal}, {Donalek}, {Larson}, \& {Christensen}}]{gra+15}
{Graham}, M.~J., {Djorgovski}, S.~G., {Stern}, D., {et~al.} 2015, \nat, 518, 74

\bibitem[{{Guo} {et~al.}(2022){Guo}, {Barth}, \& {Wang}}]{gbw+22}
{Guo}, H., {Barth}, A.~J., \& {Wang}, S. 2022, \apj, 940, 20

\bibitem[{{Guo} {et~al.}(2018){Guo}, {Shen}, \& {Wang}}]{gsw18}
{Guo}, H., {Shen}, Y., \& {Wang}, S. 2018, {PyQSOFit: Python code to fit the
  spectrum of quasars}, Astrophysics Source Code Library, record ascl:1809.008,
  , , ascl:1809.008

\bibitem[{{Gupta} {et~al.}(2018){Gupta}, {Tripathi}, {Wiita}, {Gu}, {Bambi}, \&
  {Ho}}]{gup+18}
{Gupta}, A.~C., {Tripathi}, A., {Wiita}, P.~J., {et~al.} 2018, \aap, 616, L6

\bibitem[{{Ingram} {et~al.}(2009){Ingram}, {Done}, \& {Fragile}}]{idf09}
{Ingram}, A., {Done}, C., \& {Fragile}, P.~C. 2009, \mnras, 397, L101

\bibitem[{{Ingram} \& {van der Klis}(2015)}]{iv15}
{Ingram}, A., \& {van der Klis}, M. 2015, \mnras, 446, 3516

\bibitem[{{Ingram} \& {Motta}(2019)}]{im19}
{Ingram}, A.~R., \& {Motta}, S.~E. 2019, \nar, 85, 101524

\bibitem[{{Kalberla} {et~al.}(2005){Kalberla}, {Burton}, {Hartmann}, {Arnal},
  {Bajaja}, {Morras}, \& {P{\"o}ppel}}]{kbh+05}
{Kalberla}, P.~M.~W., {Burton}, W.~B., {Hartmann}, D., {et~al.} 2005, \aap,
  440, 775

\bibitem[{{King} {et~al.}(2013){King}, {Hovatta}, {Max-Moerbeck}, {Meier},
  {Pearson}, {Readhead}, {Reeves}, {Richards}, \& {Shepherd}}]{khm+13}
{King}, O.~G., {Hovatta}, T., {Max-Moerbeck}, W., {et~al.} 2013, \mnras, 436,
  L114

\bibitem[{{Liska} {et~al.}(2018){Liska}, {Hesp}, {Tchekhovskoy}, {Ingram}, {van
  der Klis}, \& {Markoff}}]{lht+18}
{Liska}, M., {Hesp}, C., {Tchekhovskoy}, A., {et~al.} 2018, \mnras, 474, L81

\bibitem[{{Lohfink} {et~al.}(2013){Lohfink}, {Reynolds}, {Jorstad}, {Marscher},
  {Miller}, {Aller}, {Aller}, {Brenneman}, {Fabian}, {Miller}, {Mushotzky},
  {Nowak}, \& {Tombesi}}]{lrj+13}
{Lohfink}, A.~M., {Reynolds}, C.~S., {Jorstad}, S.~G., {et~al.} 2013, \apj,
  772, 83

\bibitem[{{Lomb}(1976)}]{lomb76}
{Lomb}, N.~R. 1976, \apss, 39, 447

\bibitem[{{Mainzer} {et~al.}(2011){Mainzer}, {Bauer}, {Grav}, {Masiero},
  {Cutri}, {Dailey}, {Eisenhardt}, {McMillan}, {Wright}, {Walker}, {Jedicke},
  {Spahr}, {Tholen}, {Alles}, {Beck}, {Brandenburg}, {Conrow}, {Evans},
  {Fowler}, {Jarrett}, {Marsh}, {Masci}, {McCallon}, {Wheelock}, {Wittman},
  {Wyatt}, {DeBaun}, {Elliott}, {Elsbury}, {Gautier}, {Gomillion}, {Leisawitz},
  {Maleszewski}, {Micheli}, \& {Wilkins}}]{Mainzer+11}
{Mainzer}, A., {Bauer}, J., {Grav}, T., {et~al.} 2011, \apj, 731, 53

\bibitem[{{Otero-Santos} {et~al.}(2020){Otero-Santos}, {Acosta-Pulido},
  {Becerra Gonz{\'a}lez}, {Raiteri}, {Larionov}, {Pe{\~n}il}, {Smith},
  {Ballester Niebla}, {Borman}, {Carnerero}, {Castro Segura}, {Grishina},
  {Kopatskaya}, {Larionova}, {Morozova}, {Nikiforova}, {Savchenko},
  {Troitskaya}, {Troitsky}, {Vasilyev}, \& {Villata}}]{ote+20}
{Otero-Santos}, J., {Acosta-Pulido}, J.~A., {Becerra Gonz{\'a}lez}, J.,
  {et~al.} 2020, \mnras, 492, 5524

\bibitem[{{Pan} {et~al.}(2016){Pan}, {Yuan}, {Yao}, {Zhou}, {Liu}, {Zhou}, \&
  {Zhang}}]{pan+16}
{Pan}, H.-W., {Yuan}, W., {Yao}, S., {et~al.} 2016, \apjl, 819, L19

\bibitem[{{Papoutsis} {et~al.}(2024){Papoutsis}, {Papadakis}, {Panagiotou},
  {Dov{\v{c}}iak}, \& {Kammoun}}]{ppp+24}
{Papoutsis}, M., {Papadakis}, I.~E., {Panagiotou}, C., {Dov{\v{c}}iak}, M., \&
  {Kammoun}, E. 2024, \aap, 691, A60

\bibitem[{{P{\^a}ris} {et~al.}(2017){P{\^a}ris}, {Petitjean}, {Ross}, {Myers},
  {Aubourg}, {Streblyanska}, {Bailey}, {Armengaud}, {Palanque-Delabrouille},
  {Y{\`e}che}, {Hamann}, {Strauss}, {Albareti}, {Bovy}, {Bizyaev}, {Niel
  Brandt}, {Brusa}, {Buchner}, {Comparat}, {Croft}, {Dwelly}, {Fan},
  {Font-Ribera}, {Ge}, {Georgakakis}, {Hall}, {Jiang}, {Kinemuchi},
  {Malanushenko}, {Malanushenko}, {McMahon}, {Menzel}, {Merloni}, {Nandra},
  {Noterdaeme}, {Oravetz}, {Pan}, {Pieri}, {Prada}, {Salvato}, {Schlegel},
  {Schneider}, {Simmons}, {Viel}, {Weinberg}, \& {Zhu}}]{sdssdr12}
{P{\^a}ris}, I., {Petitjean}, P., {Ross}, N.~P., {et~al.} 2017, \aap, 597, A79

\bibitem[{{Pasham} {et~al.}(2024){Pasham}, {Zaja{\v{c}}ek}, {Nixon},
  {Coughlin}, {{\'S}niegowska}, {Janiuk}, {Czerny}, {Wevers}, {Guolo}, {Ajay},
  \& {Loewenstein}}]{pas+24}
{Pasham}, D.~R., {Zaja{\v{c}}ek}, M., {Nixon}, C.~J., {et~al.} 2024, \nat, 630,
  325

\bibitem[{{Piotrovich} {et~al.}(2025){Piotrovich}, {Buliga}, \&
  {Natsvlishvili}}]{pbn25}
{Piotrovich}, M.~Y., {Buliga}, S.~D., \& {Natsvlishvili}, T.~M. 2025, Research
  in Astronomy and Astrophysics, 25, 095007

\bibitem[{{Planck Collaboration} {et~al.}(2020){Planck Collaboration},
  {Aghanim}, {Akrami}, {Ashdown}, {Aumont}, {Baccigalupi}, {Ballardini},
  {Banday}, {Barreiro}, {Bartolo}, {Basak}, {Battye}, {Benabed}, {Bernard},
  {Bersanelli}, {Bielewicz}, {Bock}, {Bond}, {Borrill}, {Bouchet}, {Boulanger},
  {Bucher}, {Burigana}, {Butler}, {Calabrese}, {Cardoso}, {Carron},
  {Challinor}, {Chiang}, {Chluba}, {Colombo}, {Combet}, {Contreras}, {Crill},
  {Cuttaia}, {de Bernardis}, {de Zotti}, {Delabrouille}, {Delouis}, {Di
  Valentino}, {Diego}, {Dor{\'e}}, {Douspis}, {Ducout}, {Dupac}, {Dusini},
  {Efstathiou}, {Elsner}, {En{\ss}lin}, {Eriksen}, {Fantaye}, {Farhang},
  {Fergusson}, {Fernandez-Cobos}, {Finelli}, {Forastieri}, {Frailis},
  {Fraisse}, {Franceschi}, {Frolov}, {Galeotta}, {Galli}, {Ganga},
  {G{\'e}nova-Santos}, {Gerbino}, {Ghosh}, {Gonz{\'a}lez-Nuevo}, {G{\'o}rski},
  {Gratton}, {Gruppuso}, {Gudmundsson}, {Hamann}, {Handley}, {Hansen},
  {Herranz}, {Hildebrandt}, {Hivon}, {Huang}, {Jaffe}, {Jones}, {Karakci},
  {Keih{\"a}nen}, {Keskitalo}, {Kiiveri}, {Kim}, {Kisner}, {Knox},
  {Krachmalnicoff}, {Kunz}, {Kurki-Suonio}, {Lagache}, {Lamarre}, {Lasenby},
  {Lattanzi}, {Lawrence}, {Le Jeune}, {Lemos}, {Lesgourgues}, {Levrier},
  {Lewis}, {Liguori}, {Lilje}, {Lilley}, {Lindholm}, {L{\'o}pez-Caniego},
  {Lubin}, {Ma}, {Mac{\'\i}as-P{\'e}rez}, {Maggio}, {Maino}, {Mandolesi},
  {Mangilli}, {Marcos-Caballero}, {Maris}, {Martin}, {Martinelli},
  {Mart{\'\i}nez-Gonz{\'a}lez}, {Matarrese}, {Mauri}, {McEwen}, {Meinhold},
  {Melchiorri}, {Mennella}, {Migliaccio}, {Millea}, {Mitra},
  {Miville-Desch{\^e}nes}, {Molinari}, {Montier}, {Morgante}, {Moss}, {Natoli},
  {N{\o}rgaard-Nielsen}, {Pagano}, {Paoletti}, {Partridge}, {Patanchon},
  {Peiris}, {Perrotta}, {Pettorino}, {Piacentini}, {Polastri}, {Polenta},
  {Puget}, {Rachen}, {Reinecke}, {Remazeilles}, {Renzi}, {Rocha}, {Rosset},
  {Roudier}, {Rubi{\~n}o-Mart{\'\i}n}, {Ruiz-Granados}, {Salvati}, {Sandri},
  {Savelainen}, {Scott}, {Shellard}, {Sirignano}, {Sirri}, {Spencer},
  {Sunyaev}, {Suur-Uski}, {Tauber}, {Tavagnacco}, {Tenti}, {Toffolatti},
  {Tomasi}, {Trombetti}, {Valenziano}, {Valiviita}, {Van Tent}, {Vibert},
  {Vielva}, {Villa}, {Vittorio}, {Wandelt}, {Wehus}, {White}, {White},
  {Zacchei}, \& {Zonca}}]{paa+18}
{Planck Collaboration}, {Aghanim}, N., {Akrami}, Y., {et~al.} 2020, \aap, 641,
  A6

\bibitem[{{Poole} {et~al.}(2008){Poole}, {Breeveld}, {Page}, {Landsman},
  {Holland}, {Roming}, {Kuin}, {Brown}, {Gronwall}, {Hunsberger}, {Koch},
  {Mason}, {Schady}, {vanden Berk}, {Blustin}, {Boyd}, {Broos}, {Carter},
  {Chester}, {Cucchiara}, {Hancock}, {Huckle}, {Immler}, {Ivanushkina},
  {Kennedy}, {Marshall}, {Morgan}, {Pandey}, {de Pasquale}, {Smith}, \&
  {Still}}]{poo+08}
{Poole}, T.~S., {Breeveld}, A.~A., {Page}, M.~J., {et~al.} 2008, \mnras, 383,
  627

\bibitem[{{Prince} {et~al.}(2025){Prince}, {Hern{\'a}ndez Santisteban},
  {Horne}, {Gelbord}, {McHardy}, {Edelson}, {Onken}, {Donnan}, {Vestergaard},
  {Kaspi}, {Winkler}, {Cackett}, {Landt}, {Barth}, {Treu}, {Valenti}, {Lira},
  {Chelouche}, {Romero Colmenero}, {Goad}, {Gonzalez-Buitrago}, {Kara}, \&
  {Villforth}}]{phh+25}
{Prince}, R., {Hern{\'a}ndez Santisteban}, J.~V., {Horne}, K., {et~al.} 2025,
  \mnras, 541, 642

\bibitem[{{Rana} \& {Mangalam}(2020)}]{rm20}
{Rana}, P., \& {Mangalam}, A. 2020, Galaxies, 8, 67

\bibitem[{{Remillard} \& {McClintock}(2006)}]{rm06}
{Remillard}, R.~A., \& {McClintock}, J.~E. 2006, \araa, 44, 49

\bibitem[{{Ren} {et~al.}(2024){Ren}, {Sun}, \& {Zhang}}]{rsz24}
{Ren}, C., {Sun}, S., \& {Zhang}, P. 2024, \apj, 961, 38

\bibitem[{{Richards} {et~al.}(2006){Richards}, {Lacy}, {Storrie-Lombardi},
  {Hall}, {Gallagher}, {Hines}, {Fan}, {Papovich}, {Vanden Berk}, {Trammell},
  {Schneider}, {Vestergaard}, {York}, {Jester}, {Anderson}, {Budav{\'a}ri}, \&
  {Szalay}}]{rls+06}
{Richards}, G.~T., {Lacy}, M., {Storrie-Lombardi}, L.~J., {et~al.} 2006, \apjs,
  166, 470

\bibitem[{{Roming} {et~al.}(2005){Roming}, {Kennedy}, {Mason}, {Nousek}, {Ahr},
  {Bingham}, {Broos}, {Carter}, {Hancock}, {Huckle}, {Hunsberger}, {Kawakami},
  {Killough}, {Koch}, {McLelland}, {Smith}, {Smith}, {Soto}, {Boyd},
  {Breeveld}, {Holland}, {Ivanushkina}, {Pryzby}, {Still}, \& {Stock}}]{rkm+05}
{Roming}, P. W.~A., {Kennedy}, T.~E., {Mason}, K.~O., {et~al.} 2005, \ssr, 120,
  95

\bibitem[{{Scargle}(1982)}]{scargle82}
{Scargle}, J.~D. 1982, \apj, 263, 835

\bibitem[{{Smith} {et~al.}(2020){Smith}, {Robles}, \& {Perlman}}]{srp20}
{Smith}, E., {Robles}, R., \& {Perlman}, E. 2020, \apj, 902, 65

\bibitem[{{Smith} {et~al.}(2018){Smith}, {Mushotzky}, {Boyd}, \&
  {Wagoner}}]{smb+18}
{Smith}, K.~L., {Mushotzky}, R.~F., {Boyd}, P.~T., \& {Wagoner}, R.~V. 2018,
  \apjl, 860, L10

\bibitem[{{Stella} \& {Vietri}(1998)}]{sv98}
{Stella}, L., \& {Vietri}, M. 1998, \apjl, 492, L59

\bibitem[{{Sun} {et~al.}(2018){Sun}, {Grier}, \& {Peterson}}]{pyccf18}
{Sun}, M., {Grier}, C.~J., \& {Peterson}, B.~M. 2018, {PyCCF: Python Cross
  Correlation Function for reverberation mapping studies}, Astrophysics Source
  Code Library, record ascl:1805.032, ,

\bibitem[{{Tarnopolski} \& {Marchenko}(2021)}]{tm21}
{Tarnopolski}, M., \& {Marchenko}, V. 2021, \apj, 911, 20

\bibitem[{{Tarnopolski} {et~al.}(2020){Tarnopolski}, {{\.Z}ywucka},
  {Marchenko}, \& {Pascual-Granado}}]{tzm+20}
{Tarnopolski}, M., {{\.Z}ywucka}, N., {Marchenko}, V., \& {Pascual-Granado}, J.
  2020, \apjs, 250, 1

\bibitem[{{Tie} \& {Kochanek}(2018)}]{tk18}
{Tie}, S.~S., \& {Kochanek}, C.~S. 2018, \mnras, 473, 80

\bibitem[{{Tonry} {et~al.}(2018){Tonry}, {Denneau}, {Heinze}, {Stalder},
  {Smith}, {Smartt}, {Stubbs}, {Weiland}, \& {Rest}}]{tdh+18}
{Tonry}, J.~L., {Denneau}, L., {Heinze}, A.~N., {et~al.} 2018, \pasp, 130,
  064505

\bibitem[{{Torrence} \& {Compo}(1998)}]{tc98}
{Torrence}, C., \& {Compo}, G.~P. 1998, Bulletin of the American Meteorological
  Society, 79, 61

\bibitem[{{Tripathi} {et~al.}(2024){Tripathi}, {Gupta}, {Smith}, {Wiita},
  {Aller}, {Volvach}, {L{\"a}hteenm{\"a}ki}, {Aller}, {Tornikoski}, \&
  {Volvach}}]{tri+24}
{Tripathi}, A., {Gupta}, A.~C., {Smith}, K.~L., {et~al.} 2024, \apj, 977, 166

\bibitem[{{Valtonen} {et~al.}(2008){Valtonen}, {Lehto}, {Nilsson}, {Heidt},
  {Takalo}, {Sillanp{\"a}{\"a}}, {Villforth}, {Kidger}, {Poyner}, {Pursimo},
  {Zola}, {Wu}, {Zhou}, {Sadakane}, {Drozdz}, {Koziel}, {Marchev}, {Ogloza},
  {Porowski}, {Siwak}, {Stachowski}, {Winiarski}, {Hentunen}, {Nissinen},
  {Liakos}, \& {Dogru}}]{val+08}
{Valtonen}, M.~J., {Lehto}, H.~J., {Nilsson}, K., {et~al.} 2008, \nat, 452, 851

\bibitem[{{Vestergaard} \& {Peterson}(2006)}]{vp06}
{Vestergaard}, M., \& {Peterson}, B.~M. 2006, \apj, 641, 689

\bibitem[{{Wang} {et~al.}(2019){Wang}, {Bai}, {Fan}, {Mao}, {Chang}, {Xin},
  {Zhang}, {Lun}, {Wang}, {Zhang}, {Ying}, {Lu}, {Wang}, {Ji}, {Xiong}, {Yu},
  {Ding}, {Ye}, {Xing}, {Yi}, {Xu}, {Zheng}, {Feng}, {He}, {Wang}, {Liu},
  {Chen}, {Xu}, {Qin}, {Zhang}, {Tan}, {Li}, {Lou}, {Li}, \& {Liu}}]{wbf+19}
{Wang}, C.-J., {Bai}, J.-M., {Fan}, Y.-F., {et~al.} 2019, Research in Astronomy
  and Astrophysics, 19, 149

\bibitem[{{Wilkins}(1972)}]{wil72}
{Wilkins}, D.~C. 1972, \prd, 5, 814

\bibitem[{{WISE Team}(2020)}]{neowise}
{WISE Team}. 2020, NEOWISE 2-Band Post-Cryo Single Exposure (L1b) Source Table,
   IPAC, doi:10.26131/IRSA124

\bibitem[{{Yu}(2002)}]{yu02}
{Yu}, Q. 2002, \mnras, 331, 935

\bibitem[{{Zhang} \& {Wang}(2021)}]{zw21}
{Zhang}, P., \& {Wang}, Z. 2021, \apj, 914, 1

\bibitem[{{Zhang} {et~al.}(2022){Zhang}, {Wang}, {Gurwell}, \&
  {Wiita}}]{zha+22}
{Zhang}, P., {Wang}, Z., {Gurwell}, M., \& {Wiita}, P.~J. 2022, \apj, 925, 207

\bibitem[{{Zhu} {et~al.}(2024){Zhu}, {Li}, {Wang}, \& {Zhang}}]{zhu+24}
{Zhu}, L.-T., {Li}, J., {Wang}, Z., \& {Zhang}, J.-J. 2024, \mnras, 530, 3538

\bibitem[{{{\.Z}ywucka} {et~al.}(2020){{\.Z}ywucka}, {Tarnopolski},
  {B{\"o}ttcher}, {Stawarz}, \& {Marchenko}}]{ztb+20}
{{\.Z}ywucka}, N., {Tarnopolski}, M., {B{\"o}ttcher}, M., {Stawarz}, {\L}., \&
  {Marchenko}, V. 2020, \apj, 888, 107

\end{thebibliography}
\bibliographystyle{aasjournal}
\end{document}